\documentstyle[aps,prb,preprint]{revtex}
\input amssym.def
\input amssym.tex
\begin{document}
\title
{On the Fermi Liquid to Polaron Crossover I: \\
General Results}
\author{A. J. Millis, R. Mueller and Boris I. Shraiman}
\address{Bell Laboratories \\
Lucent Technologies\\
600 Mountain Avenue  \\
Murray Hill, NJ 07974 }
\maketitle
\begin{abstract}
We use analytic techniques and 
the dynamical mean field method 
to study the crossover from fermi liquid to polaron
behavior in models of electrons interacting with dispersionless 
classical phonons.
\end{abstract}
\newpage

\section{Introduction}
In this paper we present the results of a study
of the crossover from fermi liquid to polaron behavior in
several related
models of electrons interacting with dispersionless classical
phonons.
We use analytic techniques valid in weak and strong
coupling limits, and
we use the
``dynamical mean field'' method \cite{Georges96}
to obtain results at intermediate couplings.
In a companion paper we present a detailed study
of a more complicated model believed to be
relevant to the "colossal magnetoresistance" manganites.

The electron-phonon problem has been extensively studied.
The ``polaron problem'' of a single electron coupled to a 
deformable medium has been understood in
detail\cite{Polaron}.
The problem of a degenerate fermi gas coupled to
phonons has been solved perturbatively in a weak
coupling limit\cite{Migdal58}.
There has also been fairly extensive work aimed at going
beyond the original perturbative solution\cite{Scalapino88,Freericks93,Freericks94,Grimaldi95},
but this work has been aimed mostly at understanding
superconducting transition temperatures and charge density
wave instabilities.
The self-trapping or polaron physics has received less attention.

There are several experimental motivations for our work.
One is the ``colossal magnetoresistance'' materials
$Re_{1-x}A_xMnO_3$\cite{Wollan55,Jin94}, where $Re$ is a rare
earth such as lanthanum and $A$ is a divalent metal ion
such as 
$Ca$ or $Sr$.  For
$0.2 \lesssim x \lesssim 0.5$ these are metals at low temperatures
and ``insulators'' (in the sense that the resistivity is high, 
and rises as $T$ is lowered) at high temperatures.
The insulating behavior has been argued to be due
to a self trapping of carriers in
local lattice distortions\cite{Millis95a,Millis95b}.
The physics of this material is complicated by the
existence of an additional ``double-exchange'' carrier
spin interaction, and will be discussed in a companion paper.
$La_{2-x}Sr_xNiO_4$ is another compound
in which carrier localization by lattice distortions has
been discussed\cite{Zaanen94}.
Previous work on $La_{2-x}Sr_xNiO_4$  has focussed on 
density wave order at particular, commensurate, $x$ but
the effects discussed here are likely to be relevant also.
We also mention the many semiconducting compounds
which display polaron effects when
very lightly doped \cite{Polaron}.
At higher dopings, fermion degeneracy effects will become
important, but the electron-phonon interaction will remain strong.

We study models of electrons interacting with phonons.
The electrons may or may not have spin and orbital degeneracy
and are coupled to one or two independent local oscillators.
The motivation for studying these different but closely related models is
as follows.
The interplay of polaron and fermi liquid physics is controlled
by energy and entropy.
The competing energies are the electron
delocalization or kinetic energy $E_{kin}$
and the lattice energy $E_{latt}$ gained by localizing
an electron.
The entropy of a localized electron depends upon the number
of spin and orbital states available for it to localize into,
while the entropy of the phonons depends on the phase space
of the oscillators.
Because the different models have different
entropies, we expect them to behave differently
at nonzero temperatures.

Our study has several limitations.
First, the numerical results are obtained via the dynamical
mean field approximation.
This is a controlled approximation which however becomes 
exact only in a
limit in which the spatial dimension $d \rightarrow \infty$
\cite{Georges96}.
Studies of other models have indicated that for $d=3$ the results
are qualitatively correct and indeed in good numerical
agreement with those obtained by other methods and by
experiment\cite{Georges96,Thomas94}.
The quantitative accuracy for quasi 2d materials
such as $La_{2-x}Sr_xNiO_4$ is not clear.
Further, the dynamical mean field approximation is essentially
local.
The only density wave instability which may be studied is the
commensurate ($\pi , \; \pi , \; \pi , \ldots$) density wave,
and the reliability of the approximation even for this
case has not been established.
We have not considered density wave solutions at
all in this work.

We have also assumed classical phonons.
This approximation is known in the single-electron case
not to affect the physics significantly \cite{Polaron} (although it does
lead to a few easily diagnosed low temperature pathologies),
and it dramatically simplifies the computations.
However, it does mean that we are unable to study superconductivity.

A third limitation is that we study models in which
the only interaction is the electron-phonon interaction.
In particular, we omit the on-site ``Hubbard-U''
Coulomb interaction, which is undoubtedly important in at least some
oxide materials\cite{Zaanen94,Tokura94}.
While the approximations we have made limit the direct
applicability of some of our results to experimental data
on these materials,
many of our results are experimentally relevant and
we believe it is useful to have a detailed account of
the behavior of a well-defined model in the literature.
At various places in the text we will qualitatively discuss
the effects of non-classical phonons and of the on-site Coulomb
interactions, but a quantitative treatment will be left to a
future publication.

The remainder of this paper is organized as follows.
Section II defines the models, the parameters, and the approximations.
Section III discusses the numerical methods used.
Section IV presents the results of a detailed solution
of the simplest model, a non-degenerate band of spinless electrons
interacting with a single oscillator.
We give phase diagrams, electron spectral functions,
frequency, and temperature dependent
conductivities and phonon probability distributions.
Section V discusses the changes that occur if electron
spin and orbital degeneracy, and multiple oscillators are included.
Section VI is a conclusion.
Appendix A clarifies the relation between the results presented here
and those of the conventional Migdal-Eliashberg treatment of the
electron-phonon problem.
The other Appendices present details of various calculations.

\section{Models and Formalism}
\subsection{Hamiltonian}
In this section we define the models we  study and present
the approximations we use.
We consider electrons coupled to phonons; the Hamiltonian, $H$,
may be written as the sum of an electronic part,
$H_{el}$, a phonon part, $H_{ph}$, and an interaction
part $H_{el-ph}$ as
\begin{equation}
H = H_{el} + H_{ph} + H_{el-ph}.
\label{eq:H}
\end{equation}
For $H_{el}$ we take electrons moving on a lattice.
An electron on a given site $i$ may be in one of $n_{orbital}$
orbital states (we will consider $n_{orbital} =1$ and
$n_{orbital} =2$) and one of $n_{spin}$ spin states (we will
consider $n_{spin} =1$ and $n_{spin} =2$).
The operator creating an electron of spin $\alpha$ in
orbital state $a$ on site $i$ is $d_{ia\alpha}^{\dagger}$ and
\begin{equation}
H_{el}= - \sum_{ij} t_{ij}^{ab}
d_{ia\alpha}^+ d_{jb\beta}
- \mu \sum_{ia\alpha}
d_{ia\alpha}^+ d_{ia\alpha}.
\label{eq:H_{el}}
\end{equation}
Here we have introduced the hopping matrix element
$t_{ij}^{ab}$ and the chemical potential $\mu$.

We model the phonons as localized classical oscillators.
We write the displacement of the oscillator
from some reference position as $r$.
We consider two cases: ``one-dimensional phonons'',
where $r$ is a scalar quantity with
$- \infty < r < \infty$, and ``two dimensional phonons''
where $\vec{r}$ is a two-component vector 
which we parametrize by a
magnitude, $r$, and an angle, $\phi$,
via $\vec{r} = r( cos \phi ,\; sin \phi )$.
We assume that in the absence of the electron-phonon
interaction the equilibrium position of the oscillator is
$r=0$, and that there is an elastic restoring force with force
constant $k$.
We measure $r$ in units of lattice constant so $r$ is dimensionless
and $k$ has the dimension of energy.  We write:
\begin{equation}
H_{ph} = \frac{1}{2} k r^2.
\label{eq:H_{ph}}
\end{equation}

The form of the electron-phonon coupling depends on the
physics.
One possible coupling is of the phonon to the charge of the
electron, i.e.
\begin{equation}
H_{el-ph}^{charge} = g \sum_{ia\alpha}
r_i (d_{ia\alpha}^+ d_{ia\alpha}
- n).
\label{eq:H_{el-ph}^{charge}}
\end{equation}
Here we have introduced the mean density
$n= \frac{1}{N} \langle \sum_{ia\alpha} 
d_{ia\alpha}^+ d_{ia\alpha} \rangle$
and have defined $r=0$ to be the equilibrium phonon state for a
uniform distribution of electrons.
The electron-phonon coupling is $g$.
If both the electrons and the phonons
are degenerate, then one may have a coupling between the orbital
state of the electron and the direction of the local phonon.
Such coupling may be expected to occur in materials such as
$La_{1-x}A_xMnO_3$ in which the Jahn-Teller effect is important
\cite{Millis95a,Goodenough55,Kanamori61}.
In such materials there are in the absence of lattice distortions
two degenerate electronic states on each site,
transforming as a two dimensional representation
of the cubic group.  The degeneracy may be split by interaction
with an appropriate phonon mode.  The interaction expressing
this physics is prescribed
by group theory to be
\begin{equation}
H_{el-ph}^{JT} = g \sum_{iab\alpha}
d_{ia\alpha}^+ \vec{\tau}^{ab}
d_{ib\alpha} \cdot \vec{r}_i.
\label{eq:H_{el-ph}^{JT}}
\end{equation}
Here $\vec{\tau} = ( \tau_z , \tau_x )$ is a vector of Pauli
matrices acting in the orbital subspace and
$r_z = r cos \phi$,
$r_x = rsin \phi$.
\subsection{Parameters and Limits}
One important energy scale is the electron
banding or kinetic energy.
This depends upon the band filling $n$ and the details of the
hopping matrix elements $t_{ij}^{ab}$, but it is roughly
of order $t$, where $t$ is a typical value of $t_{ij}^{ab}$.

Another important energy is that obtained by localizing
an electron on a site.
By minimizing Eqs. \ref{eq:H_{el-ph}^{charge}} or \ref{eq:H_{el-ph}^{JT}}
and \ref{eq:H_{ph}} one
finds this energy to be of order $g^2/k$.
The ratio of these scales is the important dimensionless
parameter $\lambda = g^2 /kt$; $\lambda < 1$ corresponds
to weak coupling and $\lambda > 1$  to strong coupling.
This $\lambda$ is shown in Appendix A to be related to the
usual coupling constant of the Migdal-Eliashberg
theory by a factor of $n_{spin}n_{orb}/\pi$.

A final important scale is the phonon frequency $\omega_D$.
In the weak electron-phonon coupling limit it is known
that for $T \gtrsim \omega_D/3$, phonons behave
classically while for lower $T$ quantum effects are important
\cite{Polaron}.
In the strong coupling limit, polarons are formed and large lattice
distortions occur. From studies of a single 
electron coupled to a deformable medium \cite{Polaron}
one finds that quantum effects are important only at a very
low scale $\omega_D/n$, where $n \gg 1$ is roughly the number of
quanta involved in the formation of the polaron \cite{Polaron}.
Throughout this paper we assume that $T$ is sufficiently high that quantum
effects in the phonons may be neglected.
\subsection{Formalism}
Physical quantities may be calculated from the partition
function, $Z$, which may be written as a functional integral over
the electron and phonon fields.
Because we have assumed classical phonons, we may solve the electron
problem for given values of the phonon coordinate and then average
over the phonon coordinates.
Because the phonon is classical, we may rescale the phonon
coordinate to $r = \sqrt{k/t} \; r$.
The partition function becomes
\begin{equation}
Z = \int {\cal D}_{phonon} \;
exp \left [ - \frac{1}{2} \frac{t}{T} \sum_i r_i^2 + 
Tr \; ln \; \left [ G_{eff}^{-1} \right ] \right ].
\label{eq:Z}
\end{equation}
Here $T$ is the temperature,
${\cal D}_{phonon}$ is the phonon measure appropriate
to the choice of phonon and $G_{eff}$ is the electron Green function
appropriate to Eq. \ref{eq:H} with fixed static $r_i$.
Note also that $r$ has been rescaled in such a way that
in $H_{el-ph}$ the coupling
constant $g = \sqrt{\lambda}$.

Starting from Eq. \ref{eq:Z} one may make perturbative diagrammatic
calculations using standard methods.
In the weak coupling limit the starting point is
$r_i=0$ (corresponding to uncoupled electrons and phonons) and it
may be seen from the first term in the exponent of
Eq. \ref{eq:Z} that fluctuations
in $r$ are small, of order $\sqrt{T/t}$.
This is  the classical analogue of the Migdal's
expansion parameter $\sqrt{\omega_D/t}$.
In the strong coupling limit the starting point has some
$r_i \neq 0$ but fluctuations are still of order
$\sqrt{T}$, so that at low $T$ an analytic treatment is 
possible, if the ground-state configuration
of the $r_i$ can be determined.
The connection to the conventional treatment of the electron-phonon
problem is discussed in more detail in Appendix A.

At intermediate coupling we have not succeeded in constructing 
the ground state about which to expand.
We have therefore resorted to the ``dynamical mean field''
approximation\cite{Georges96} in which it is assumed
that the electron self energy is local (i.e. momentum independent).
In the electron-phonon problem in three spatial dimensions 
the self energy is known
to have only a weak momentum dependence in the various solvable
limits\cite{Migdal58,Abrikosov63},
so the dynamical mean field approximation, 
which formally becomes exact in a limit in which
spatial dimensionality $d \rightarrow \infty$, seems likely to be reliable.

The electron Green function $G$ in general depends upon
orbital indices (a,b), spin indices ($\alpha , \beta $),
momentum $\vec{k}$ and Matsubara frequency $i \omega_n$.
The dynamical mean field {\it Ansatz} is
\begin{equation}
G_{\alpha\beta}^{ab} (k, i \omega_n )=
\left [ i\omega_n- \Sigma_{\alpha \beta}^{ab} (i\omega_n)
- \epsilon_k^{ab} + \mu \right ]^{-1}.
\label{eq:G_{alpha beta}^{ab}}
\end{equation}
Here $\epsilon_k^{ab}$ is the Fourier transform of $t_{ij}^{ab}$.
All of the quantities in
Eq. \ref{eq:G_{alpha beta}^{ab}} are
tensors in the direct product of spin and orbital space,
and quantities without Roman ($a,b$) or Greek ($\alpha , \beta $) indices
are proportional to the unit matrix in orbital and spin space respectively.

Throughout this paper we shall assume that there is no long
range order in either orbital space or spin space,
so we may take
$\Sigma_{\alpha\beta}^{ab} \sim \delta_{ab} \delta_{\alpha \beta}$.
Generalizing the formalism to include orbital ordering
would be straightforward and quite interesting.  The
generalization to ferromagnetic spin ordering is
given in a companion paper \cite{Millis96b}.
Because $\Sigma$ is taken to be $k$-independent all interaction effects
are derivable from the local ($k$-integrated) Green function,
$G_{LOC}$.  Because we have assumed that there is no long-range
order in orbital space, $G_{LOC}^{ab} \sim \delta_{ab}$.
We may therefore write
\begin{equation}
G_{LOC} (i\omega_n ) =
\frac{1}{2} \; Tr \; \int
\frac{d^dk}{(2\pi )^d} \; G^{ab} (k, i\omega_n).
\label{eq:G_{LOC}Tr}
\end{equation}
The $k$ integral may be simplified by introducing
at each $k$-point the $ab$ space rotation $R_k$ which diagonalizes
$G^{ab}$, writing
$G^{ab}(k)=R_kG_{Diag}R_k^{-1}$,
and exploiting the cyclic invariance of the trace.
The $k$ integrals of the two components of $G_{Diag}$ must be identical,
so we may finally write
\begin{equation}
G_{LOC} (i\omega_n) =
\int \; \frac{d\epsilon_k {\cal D} (\epsilon_k)}
{i\omega_n - \Sigma (i\omega_n) - \mu - \epsilon_k}
\label{eq:G_{LOC}}
\end{equation}
with ${\cal D}(\epsilon_k)$ the density of states.

Because all interactions are local, $G_{LOC}$ may be derived from
a partition function identical in form to Eq. \ref{eq:Z} but
with spatial index $i$ suppressed and the
quantity $G_{eff}^{-1}$ replaced by 
\begin{equation}
G_{MF}^{-1} = - a \; d_{a\alpha}^+ d_{a\alpha} +
H_{el-ph}
\label{eq:H_{LOC}}
\end{equation}
at fixed phonon coordinate.  The free energy corresponding to
a given value of phonon coordinate may be found as usual
from $G_{MF}$, and from this one may construct the
phonon probability distribution
\begin{equation}
P(r)= \frac{1}{Z_{LOC}} exp 
\left [ - \frac{r^2}{2T} +
\sum_n Tr \; ln \; G_{MF}^{-1} \right ].
\label{eq:P}
\end{equation}
where the local-approximation partition function $Z_{LOC}$
is defined by

\begin{equation}
Z_{LOC} = 
\int {\cal D}_{phonon} P(r)
\label{eq:Z_{LOC}}
\end{equation}
Here $a$ is a frequency-dependent effective field (analogous to the
magnetization in the usual Weiss mean field method) determined 
by the condition
\begin{equation}
G_{LOC}(i\omega_n) = \frac{\partial \; ln\; Z_{LOC}}
{\partial a (i \omega_n)}.
\label{eq:{consist}}
\end{equation}
This condition may be implemented by observing that
\begin{equation}
\Sigma (i\omega_n) = a(i\omega_n) - G_{LOC}^{-1}.
\label{eq:sigma}
\end{equation}
The precise form of the resulting equation depends on the density of states.
We shall consider two forms, the Lorentzian
\begin{equation}
{\cal D}_{Lorentzian} ( \epsilon_k) =
\frac{t/ \pi}{\epsilon_k^2 + t^2}
\label{eq:D_{Lorentzian}}
\end{equation}
and the semicircular,
\begin{equation}
{\cal D}_{semi} (\epsilon_k) = 
\sqrt{4t^2-\epsilon_k^2}
/ (2 \pi \; t^2).
\label{eq:D_{semi}}
\end{equation}
The Lorentzian density of states has the advantage
that the self consistency equation may be analytically solved
to yield
\begin{equation}
a_{LOR} (i\omega_n) =
i\omega_n + \mu + i t  sgn\omega_n.
\label{eq:a_{Lor}}
\end{equation}

It has however two unphysical features:
${\cal D}_{Lor}(\epsilon_k)$ is unbounded and
$\int_{-\infty}^{\mu} d\epsilon_k \epsilon_k {\cal D}_{Lor} (\epsilon_k) 
= \infty$ so
the kinetic energy is infinite for finite $\mu$.

For the semicircular density of states the $\epsilon_k$ integral
may also be performed analytically, leading to
\begin{equation}
a_{semi} (i\omega_n) =
i\omega_n + \mu -t^2
G_{LOC} (i\omega_n).
\label{eq:a_{semi}}
\end{equation}
This equation must be solved numerically, except in
simple limits.

The semicircular density of states corresponds to a Bethe lattice
and the dynamical mean field theory together with the self-consistency
condition Eq. \ref{eq:a_{semi}} can be derived 
in this case by a more physically
transparent argument.  We take for
concreteness the case of orbitally degenerate electrons
with a Jahn-Teller coupling to orbitally degenerate classical
phonons.  On a Bethe lattice the local Green function $G_{ii}$
on a site $i$ with phonon amplitude $r_i$ is:

\begin{equation}
{G_{ii} ^{ab} }^{-1} (i \omega_n ,r_i) \ = \ i \omega_n +\mu +
g \vec{r} \cdot \vec{\tau^{ab}} - \sum_{jcd} t_{ij} ^{ac}
G_{jj} ^{cd}  (i \omega_n,r_j ) t_{ji} ^{db}
\label{eq:bethe}
\end{equation}

One may diagonalize the Jahn-Teller coupling 
by a local rotation $R_i$ in  orbital space, with $R_i =
cos \phi_i /2 \ \tau_z + sin \phi_i /2 \ \tau_x $, and
one may absorb $R_i$  into
the hopping matrix $\hat{t}_{ij} = R_i t_{ij} R_j$ which
thus acquires
a dependence on the relative orientation of the phonon distortions.
The mean field theory for the disordered phonon state
follows (in the limit of large coordination
number) from the assumption that the $\phi_j$'s and $r_j$ are statistically
independent and the sum on $j$ can be replaced by an average over $\phi$'s
and $r's$, the latter with a non-trivial distribution $P(r)$ defined by
Eq. \ref{eq:P}. The hopping self-energy then
 reduces to $\delta^{ab} G_{LOC}$
with $G_{LOC} = 1/2 \ Tr \ < G_{ii} > _{r_i}$ and the self-consistent
equation for $G_{LOC}$ is obtained by averaging $G_{ii}$ given by
Eq. \ref{eq:bethe} over $r_i$.  Eq. \ref{eq:bethe} may be
helpful as a starting point for 
mean field theories with non-trivial
spatial correlations of the phonons.

We will be interested in three physical quantities:
the momentum-integrated electron spectral function
$A ( \omega )$, the phonon probability distribution
$P(r)$, and the conductivity $\sigma$.
The spectral function is
\begin{equation}
A(\omega ) =- \frac{Im \; G_{LOC}
(i \omega_n \rightarrow \omega +i \epsilon)}{\pi}.
\label{eq:A}
\end{equation}

In the dynamical mean field method the conductivity
$\sigma_{ab}=\sigma_{MF} \delta_{ab}$ with
\begin{equation}
\sigma_{MF} (i \Omega ) =
\frac{e^2t^2}{i \Omega} Tr
\int d\epsilon_p {\cal D}
(\epsilon_p) T \sum_{i\omega}
{\bf G} (p,i\omega)
{\bf G}(p, i\omega + i \Omega ).
\label{eq:sigma_{MF}}
\end{equation}
In particular, the dc conductivity is
\begin{equation}
\sigma_{dc}(T) = e^2t^2 Tr
\int \frac{d\epsilon d\omega}{\pi}
{\cal D} (\epsilon_p)
[Im \; {\bf G}(\epsilon_p, \omega ) ]^2/4T cosh(\omega /2T)^2
\label{eq:sigma_{dc}}
\end{equation}
with ${\bf G}(\epsilon_p,\omega )^{-1} = 
\omega + \mu -(\epsilon_p^{ab}) - \Sigma (\omega )$.

As $T \rightarrow 0$, two possibilities arise:
either $\Sigma^{\prime \prime} (\omega =0) \rightarrow 0$ 
or it tends to a finite non-zero value.
In the former case,
\begin{equation}
\sigma_{dc} \rightarrow e^2t^2
n_{spin}n_{orb} {\cal D} (\epsilon^*)
\int \frac{d\omega}{4T}
\frac{1}{(cosh^2 \frac{\omega}{2T} )}
\frac{1}{\Sigma^{\prime \prime} (\omega ,T)}
\label{eq:sigma_{dc}1}
\end{equation}
with $\epsilon^*$ satisfying $\epsilon^*= \mu- \Sigma^{\prime} (0)$.
In the latter case
\begin{equation}
\sigma_{dc} \rightarrow \frac{e^2t^2}{\pi}
n_{spin}n_{orb} \int d \epsilon_p
{\cal D} ( \epsilon_p)
\left [ \frac{\Sigma^{\prime \prime}(0)}
{(\epsilon_p +\mu- \Sigma^{\prime}(0))^2+ 
\Sigma^{\prime \prime} (0)^2} \right ]^2.
\label{eq:sigma_{dc}2}
\end{equation}
\section{Numerical Methods}
In this section we outline the numerical methods
used to solve Eq. \ref{eq:a_{semi}} and compute physical
quantities.
To aid in the discussion we write this equation explicitly
for the simplest case of nondegenerate electrons coupled to
nondegenerate phonons.
We have, setting $t=1$,
\begin{equation}
a (\omega ) = i \omega + \mu -
\int_{- \infty}^{\infty} dr
\frac{P(r)}{a(\omega) + gr}
\label{eq:explicit}
\end{equation}
with
\begin{equation}
P(r) = \frac{1}{Z_{LOC}} exp
\left [ - \frac{r^2}{2T} + \sum_n \; ln \;
\left [ \frac{a_n+gr}{i \omega_n} \right ] \right ]
\label{eq:P(r)}
\end{equation}
and $Z_{LOC}$ given by Eq. \ref{eq:Z_{LOC}}.
Here $a_n=a(i\omega_n)$ with $\omega_n$ a fermion
Matsubara frequency
$\omega_n= (2n+1) \pi T$.

For most physical quantities we require $a(\omega )$ for
real frequency.
It is possible to express $P(r)$ in terms of $a(\omega )$,
but this involves performing numerically an integral over a
continuous frequency.
We have found it more convenient to find $P(r)$ by solving
Eq. \ref{eq:explicit} on the Matsubara points and then
to use this $P(r)$ to solve for $a(\omega)$ on the real axis.
In what follows we first discuss the solution on the Matsubara
axis and then mention the additional issues arising for the real
axis case.

We solve Eq. \ref{eq:explicit} on the Matsubara axis
by direct iteration, so that $a_n$ at step $N+1$ is
determined by evaluating the right-hand side of
Eq. \ref{eq:explicit} using $a_n$ from step N.
If a solution at the same coupling and a nearby temperature
is available we use this as the starting point;
if not we use the $g=0$ solution.
Convergence is usually rapid; typically 10 iterations
are required for convergence from the $g=0$ solution to a
solution with RMS error of $10^{-5}$ (averaged over all
retained Matsubara frequencies) and using the solution from a
nearby $T$ saves about 2 iterations.
However away from half filling, in the strong
coupling limit, the straightforward iteration procedure
does not converge, but instead goes to a two-step limit cycle in
which evaluating the right hand side of Eq. \ref{eq:explicit} using
the values $a_n^{(1)}$ yields a different set of values
$a_n^{(2)}$ which, when put back in Eq. \ref{eq:explicit},
generate again $a_n^{(1)}$.
This limit cycle may be avoided by making the iteration proceed in
smaller steps by defining $a_n$ at step $N+1$ to be an appropriate linear
combination of the right-hand side of Eq. \ref{eq:explicit}
and the solution at step N; however as $T$ gets lower or 
coupling gets stronger the
required admixture of the newly computed $a$ gets smaller, and at some
point the computations become too time consuming.

Away from half filling it is necessary at each temperature
and coupling to find the chemical potential $\mu (T,g)$ which
gives the desired $n$.
We do this by computing at several values of $\mu$ and
interpolating.
At strong coupling and low $T$, $n$ is a very sensitive function of
$\mu$, and it is necessary to choose at least three $\mu$-values
all of which yield $n$'s which are within 10 percent
of the desired value and to use a cubic interpolation.
In the calculations presented $n$ varies by less than
$6 \times 10^{-4}$ over the range of $T$ and $g$ considered.
We found it convenient to evaluate $n$ via
\begin{equation}
n= lim_{\tau \rightarrow 0} T \sum_n Tr \;
G_{LOC} (i \omega_n) \;
e^{-i\omega_n\tau}.
\label{eq:n}
\end{equation}
$G_{LOC}$ may be expressed in terms of a via the mean field equations.

The next issue is the number of Matsubara points needed to compute
$P(r)$ accurately.
We note from Eq. \ref{eq:explicit} that at large $\omega_n$
\begin{equation}
a_n= i \omega_n+\mu -
\frac{1}{i\omega_n} -
\frac{1}{\omega_n^2} \int dr \;
P(r)
(\mu +gr) + {\it O} \left ( \frac{1}{\omega^3} \right )
\label{eq:large}
\end{equation}
Analogous but slightly different formulas apply in the case of
Jahn-Teller coupling.
We determine the maximum retained Matsubara frequency $n_{max}$
by requiring that the ${\it O}(1/ \omega_n^3)$ terms are less
than $10^{-5}$ and in the expression for $P(r)$ we
evaluate the terms with $|n | > n_{max}$ analytically, using
the $a_n$ given in Eq. \ref{eq:large};
the errors are of order $n_{max}^{-2}$.
Equation \ref{eq:large} is also useful for improving the accuracy
of the computation of the density $n$ from Eq. \ref{eq:n}
away from half-filling.
To achieve the stated accuracy we found it necessary to choose
$n_{max}$ so that $\omega_{n_{max}} = 2 \pi T(n_{max}+1/2)$ is
about five times the full width of the real-axis electron
spectral function.  The
$1/T$ growth of $n_{max}$ as
$T \rightarrow 0$ limits our ability to 
calculate at very low temperatures.
For example at $g=1.29$ and $T=0.02$, $n_{max} \cong 250$.

The remaining numerical issue for the Matsubara-axis
solution is the integral over the phonon coordinate,
which is done using the numerical quadrature routine ``quad''
from the ``Port'' library.
Gaussian quadrature routines are faster at weak coupling,
but fail at strong coupling and low $T$ where $P(r)$ has a large 
narrow
peak at an a-priori unknown location far from the origin.

We now turn to the real axis calculations, which we also do by
direct iteration of Eq. \ref{eq:explicit} with analytically
continued frequency $\omega+i\epsilon$, using $P(r)$ which has
already been computed.
The limit cycles which plagued the Matsubara calculations
do not occur, but other numerical problems arise.
Because $P(r)$ is held fixed, different
frequencies decouple and the mean field equation does not come
from minimizing a free energy.
Nearby frequencies are also sometimes found to converge
at different rates.
In parameter regimes in which the $T=0$ spectral function has a
gap, the imaginary $a^{''}(\omega )$ part of
$a(\omega )$ is small
and very temperature
dependent at low $T$ for frequencies in the gap.
Our procedure fails to find a non-zero value of $a^{''}$ if this
quantity becomes less than $10^{-3}$, and we are not sure
of the accuracy for $a^{''} \lesssim 10^{-2}$.
Physical properties in such regimes depend crucially
on the tailing of the spectral function into the gap, so our
ability to calculate the low $T$ properties of regimes with gaps is
limited.

To calculate conductivities we solve the real axis mean field
equations for a discrete set of frequencies spaced by $0.01$.
Then at these frequencies we evaluate
\begin{equation}
\theta(\omega )= \int d\epsilon_p
{\cal D} (\epsilon_p) [ Im \;
G(\epsilon_p, \omega ]^2
\label{eq:phi}
\end{equation}
using the Port routine ``quad'' and then obtain
\begin{equation}
\sigma  \simeq n_{spin} n_{orb}
\int \; d\omega \; \theta (\omega ) \;
cosh^{-2} \frac{\omega}{2T}
\label{eq:sigma-num}
\end{equation}
using the quad routine and evaluating $\theta (\omega )$ by spline
interpolation from the known values.
In the computation we set the factors
$e^2 / \pi = 1$.

We note finally that especially in the crossover region where
there is a frozen-in lattice distortion which is not large enough
to open a gap in the spectral function, physical properties are
quite sensitive to the precise value of the coupling
constant.
A change in $g$ by 2\% can lead to more than factor of three
changes in e.g. the low $T$ conductivity; also small numerical
errors in the computation of the fermion contribution to the action
can mimic a small change in $g$.
An example can be found in the curve labelled
$g=1.60$ in Fig. 2.
The analytically calculated $T=0$ resistivity for this $g$ is 3.75.
The actual numerically calculated low $T$ spectral function
is that expected for $g=1.57$; the corresponding analytical
$T=0$ resistivity is found to be $1.004$, in good agreement
with the numerical calculation.
\section{Non-degenerate Spinless Electrons}
\subsection{Introduction}
In this section we present results of a detailed study of
a model of spinless, orbitally non-degenerate fermions coupled to
classical non-degenerate oscillators.
This section is organized as follows.
In this introduction, we outline the qualitative physics.
In part 4-B we solve the original model,
Eqs. \ref{eq:H}, \ref{eq:H_{el}}, 
\ref{eq:H_{ph}}, \ref{eq:H_{el-ph}^{charge}} in
simple limits, obtaining results which clarify the interpretation of the
$d= \infty$ results.
In part 4-C we solve the $d= \infty$ model with the Lorentzian
density of states, (Eqs. \ref{eq:Z_{LOC}}, \ref{eq:a_{Lor}}),
and in part 4-D the $d= \infty$ model with semicircular density of states
(Eqs. \ref{eq:Z_{LOC}}, \ref{eq:a_{semi}}).

In the model defined by Eqs. 1-4 we may distinguish three
phases according to the $T \rightarrow 0$ limits of the
oscillator coordinate $r$ and the electron spectral function.
In {\it weak coupling}, $lim_{T\rightarrow 0} \langle r \rangle =0$
and the spectral function takes the non-interacting form.
The low $T$ resistivity is then linear in $T$, and extrapolates
to 0 as $T \rightarrow 0$, being due to the thermal fluctuations
of $r$ about $r=0$.
In {\it intermediate coupling},
$lim_{T\rightarrow 0} \langle r \rangle \neq 0$, implying
that some frozen-in lattice distortions exist.
However, the amplitude of these distortions is too small to
localize states near the fermi level.
The low-$T$ resistivity is still linear in $T$, but extrapolates to a
non-zero value at $T=0$.
The spectral function is perturbed from its non-interacting
form.
In {\it strong coupling} the amplitude of the frozen-in lattice distortions
is large enough to localize electrons at the Fermi level,
leading to a gap in the $T=0$ spectral function and an
activated resistivity.

All three regimes are found in the calculations
discussed below.
None of the regimes involves long
rang order-- the localization should be thought of as being due
to polaron formation.
At commensurate filling or in low dimensions,
effects of long-range order (i.e. charge density wave formation)
will be important but at general incommensurate fillings these
effects may be neglected.

Quantum effects will also be important:
at sufficiently low $T$ and in the absence of long range order,
quantum tunnelling
will restore translational invariance.
In the strong coupling limit the quantum effects may be
thought of as leading to the formation of a polaron band,
with a band width exponentially small in $\lambda$;
at scales greater than the polaron band width the results presented
here will apply.
The effect of quantum fluctuations on the intermediate coupling
regime is less clear, but must be left for future work.
\subsection{Full Model, Solvable Limits}
The crucial issue is performing the integral
over the phonon coordinates in
Eq. \ref{eq:Z}.
At low $T$ the integral is sharply peaked about
particular values $\{ r_i^* \}$ and a steepest descents approximation is possible.
Two solvable cases occur: \\
1. {\it Weak coupling}.
For $\lambda < \lambda_c$ and $T \rightarrow 0$
($\lambda_c$ will be estimated below)
$r_i^* \equiv 0$ and one may expand the logarithm in
Eq. \ref{eq:Z} obtaining
\begin{equation}
Z \propto exp \; \left [ Tr \; ln \; G_0^{-1} \right ]
\int {\cal D}r_k \;
exp - \left [ \frac{t\; (1 - \lambda B(k))\;r_k r_{-k}}
{2T} \right ].
\label{eq:Z_{weak}}
\end{equation}

Here $G_0^{-1} = i\omega_n -\epsilon_k+ \mu$ and
\begin{equation}
B(k) = t \int \frac{d^3p}{(2 \pi )^3}
\frac{f(\epsilon_{p+k})-f(\epsilon_p)}
{\epsilon_p - \epsilon_{p+k}}.
\label{eq:B}
\end{equation}
It is evident from Eq. \ref{eq:Z_{weak}}
that the expansion in powers of $r$ about
$r_i \cong 0$ is controlled at low $T$
if $\lambda < \lambda_c$ with
\begin{equation}
\lambda_c^{-1} = max_k \; B(k).
\label{eq:lambda_c}
\end{equation}

In particular, as $T \rightarrow 0$ at fixed
$\lambda$, $r_k \sim \sqrt{T}$ is small, so the
leading order perturbation expression for the fermion self energy,
$\Sigma$, applies  \cite{Migdal58,Abrikosov63}.
This expression is:
\begin{equation}
\Sigma (p, i\omega , T) = \lambda T \int
\frac{d^3k}{(2\pi )^3}
\frac{1}{1-\lambda B(k)}
\frac{t}{i\omega - \epsilon_{p+k}}.
\label{eq:Sigma(p,i omega ,T}
\end{equation}
For small $\omega_n$,
we may restrict the $k$ integral to wavevectors such that
$p^{\prime} = p+k$ is on the Fermi surface.
The leading contribution to the $\omega$ dependence of $\Sigma$
is then
\begin{eqnarray}
\Sigma(p,i\omega ,T) &=& i \pi T \lambda \;
sign \: i \omega_n \nonumber \\
& \int & \frac{d\Omega_p^{\prime}}{4 \pi}
{\cal D} (\epsilon_{p^{\prime}}) / [ 1- \lambda B(p-p^{\prime} )]
\label{eq:Sigma_{small}omega}
\end{eqnarray}
and if also $\lambda \ll \lambda_c$
\begin{equation}
\Sigma (p,i\omega ,T) = i \pi T \lambda
\bar{\cal D} (\epsilon_F) \; sgn \omega_n
\label{eq:Sigma_{small}omega small lambda}
\end{equation}
with $\bar{\cal D}(\epsilon_F)$ the angle averaged
density of states at the fermi surface.

Equation \ref{eq:Sigma_{small}omega small lambda} 
is shown in Appendix A to be precisely
the usual Migdal expression for the self energy.
If $( \lambda_c-\lambda ) / \lambda_c$ is not too small,
the quantity
$\lambda T \int \frac{d\Omega_{p^{\prime}}}{4\pi}$
${\cal D} (\epsilon_{p^{\prime}}) /[1-\lambda B(p-p^{\prime} )]$
depends only weakly on $p$; if it is averaged over $p$ or the Fermi
surface it becomes the classical limit of the usual ``$\alpha^2F$''
function.
Here $T/t$ plays the role of the usual Migdal parameter
$\omega_D /E_F$.
As $\lambda \rightarrow \lambda_c$, the phonon
propagator becomes large and also acquires a singular momentum
dependence because the phonon propagator diverges at a
particular wavevector, so one may expect the Migdal
approximation to fail.
A discussion of corrections to the Migdal
approximation is beyond the scope of this paper.
Note also that $\lambda < \lambda_c$ is the condition for
the linear stability of
the $r_i \equiv 0$ state.
A $T=0$ first order transition to a state of
$r_i \neq 0$ at a $\lambda^* < \lambda_c$ is possible and
indeed will be demonstrated in the dynamical mean field
approximation.

The resistivity may also be computed.
If $\lambda \approx \lambda_c$, the momentum dependence of
the phonons may be important and vertex corrections must be
considered.
If $\lambda \ll \lambda_c$ the momentum dependence of the
phonons, and therefore the vertex corrections, may be neglected
and the conductivity on the Matsubara axis is
\begin{eqnarray}
\sigma_{ab} (i\Omega_n) &=& \frac{e^2}{i\Omega_n}
\int \frac{d^3p}{(2\pi )^3}
T \sum_{i\omega_n}
\frac{\partial \epsilon_p}{\partial p_a}
\frac{\partial \epsilon_p}{\partial p_b} \nonumber \\
     & \cdot &G (p, i\omega_n+i\Omega_n) G(p,i\omega_n).
\label{eq:sigma(i Omega_n)}
\end{eqnarray}

Substituting Eq. \ref{eq:Sigma_{small}omega}
into Eq. \ref{eq:sigma(i Omega_n)}, performing
the integrals and analytically continuing gives
\begin{equation}
\sigma (T) = \frac{e^2}{2\pi \lambda T} \int
\frac{d\Omega_p}{4\pi}
(v_F^a)^2 {\cal D}(\Omega_p)
\label{eq:Sigma(p,i omega ,T)}.
\end{equation}
In a simple tight binding model,
$v_x(p)= 2t sin p_x$ and evaluating
Eq. \ref{eq:Sigma(p,i omega ,T)} we have
\begin{equation}
\rho (T) = \frac{\pi \lambda T}{t^2e^2 \bar{\cal D} (\epsilon_F)}.
\label{eq:rho(T)}
\end{equation}

\noindent
2. {\it Full Model, Strong Coupling}.
In  the limit $\lambda \gg \lambda_c$ an analytic solution is
possible because we may treat the electron kinetic
energy as a perturbation.
If it is neglected then to solve Eq. 1-4 we note that if
the mean density is $n$ there are $n$ sites with one electron,
and $1- n$ sites with no electron.
On the $n$ occupied sites there is a lattice
distortion of amplitude
\begin{equation}
r_{occ}^* = -(g/k) (1- n)
\label{eq:r_{occ}^*}
\end{equation}
and on the $1-n$ unoccupied sites
\begin{equation}
r_{unocc}^* = (g/k) n.
\label{eq:r_{unocc}^*}
\end{equation}
The energy gained is found from Eq \ref{eq:H} to be
\begin{equation}
E_{gained} = - \frac{1}{4} t \lambda ((1- n)^2 +
n^2).
\label{eq:E_{gained}}
\end{equation}
Perturbative corrections (in powers of the kinetic energy)
may be calculated.
They involve processes in which electrons make virtual
hops to unoccupied neighboring sites, and lead to a
repulsive short ranged interaction of order $t/ \lambda$
between polarons.
This interaction will lead to polaron ordering, at a scale
generically much lower than $t/ \lambda$.

In this limit transport is thermally activated; the prefactor is
$t$ and the gap is $E_{gained}$.
\subsection{Dynamical Mean Field Method: Lorentzian Density of States}
In this subsection we discuss the solution of the 
dynamical mean field equations for spinless fermions with Lorentzian
density of states coupled to a single scalar oscillator.
We begin with the $T \rightarrow 0$ limit, where one may
use a steepest descents approximation.
The details are given in Appendix
B.  We find that for $\lambda < \lambda_c(n)$, $r^* = lim_{T \rightarrow
0} \langle r \rangle =0$, while for $\lambda > \lambda_c$ there
are two extrema, at $r^*_1 <0$ and $r^*_2>0$.
For $n=1/2$, 
$r^*_1=-r^*_2$, the transition at $\lambda_c$ is second order;
and $r^*$ grows smoothly with $\lambda$.
For $n \neq 1/2$, $r^*_1 \neq -r^*_2$ and the transition is first
order.

The physical content of the $r^* \neq 0$ states is most 
clearly seen
from the electron Green function which at
$T=0$ is
\begin{equation}
G_{LOC}(i\omega ) =
\frac{Z_1}{i\omega +\mu +itsgn\omega + g r_1^*}
+ 
\frac{Z_2}{i\omega+\mu+itsgn \omega  + g r_2^*}
\label{eq:G_{LOC-LOR}}
\end{equation}
with $Z_1=\frac{|r_2^*|}{|r_1^*|+|r_2^*|}$
and $Z_2=\frac{|r_1^*|}{|r_1^*|+|r_2^*|}$
Thus if $r^* \neq 0$ the spectral function
consists of two resonances.  In the large $\lambda$ limit, these
are well separated.
One, of weight $n$, is at the negative energy
$\mu- g(1-n)$ and corresponds to the
sites on which the electron-phonon interaction
has localized an electron.
The other, of weight $1-n$, is at the positive energy
$\mu+ g n$ and corresponds to the empty sites.
The splitting, $\Delta E \propto \lambda$ as expected,
while the energy gained $E_{gained}$ (relative to the
undistorted state) is 
\begin{equation}
E_{gained}=- \frac{1}{4} \lambda t[(1-n)^2+n^2]
\label{eq:E_{gained-Lor}}
\end{equation}
consistent with Eq. \ref{eq:E_{gained}}.
The Lorentzian density of states is pathological in that,
as may be seen from Eq. \ref{eq:G_{LOC-LOR}}, at any value
of $g$ the $T=0$ spectral function is non-zero at
$\omega =\mu$; thus the ground state is always metallic.
This pathology may be traced
back to the non-integrability of the first moment of
density of states, which implies that some carriers
have arbitrarily large velocities and cannot be localized
by any lattice distortion.

\subsection{Dynamical Mean Field Method:
Semicircular Density of States}

In this subsection we present results obtained using the
dynamical mean field method with semicircular density
of states.
Our analysis is based on Eqs. \ref{eq:Z_{LOC}},\ref{eq:a_{semi}}.
The technical differences from the previously discussed
Lorentzian case are that the mean field function $a(\omega )$
satisfies a non-trivial self-consistency equation, and the
moments of the density of states are finite, so
the spectral function may develop a gap.

As in the previous subsection, one may use
steepest descents arguments for T near zero.
The details of the analysis are given in Appendix C.
For $g < g^*$ the energy minimum is
at $r^* = 0$; for $g > g^*$ there
are two minima, at $r=r_1>0$ and
$r=r_2<0$.
For $n=1/2$ the transition is second order and occurs
at $g^*=g_c^*= \sqrt {3 \pi} /2\cong 1.535$;
for $n \neq 1/2$ the transition is first order and occurs
at a $g^* > g_c$.

As in the previous section, the physical content of the
$r^*\neq0$ solutions is the existence of
frozen-in lattice distortions.
The effect of these may be seen from the local Green function which,
at $T=0$ is
\begin{equation}
G_{LOC}= \left (
\frac{Z_1}{a( \omega ) + g r_1} ~+~
\frac{Z_2}{a( \omega ) ~+~ g r_2} \right )
/ (Z_1 + Z_2)
\end{equation}
Here $Z_1>0$ and $Z_2>0$ are the
relative weights of the free energy minima corresponding
to $r=r_1$ and $r=r_2$ respectively.

Figure 1 shows $A(\omega)= -ImG_{LOC}$ for
$n=1/2$ and different values of $r_1=r_2=r$.
Our interpretation of these results is that for 
$0<r<1/2$ the
potential fluctuations due to the frozen-in phonons are very
weak, and merely localize a few states at the band tails.
This accounts for the slight broadening of A and the slight decrease
near $\omega=-\mu$.
For $1>r>1/2$ the potential is strong enough
to create a minimum in A at $\omega=-\mu$;
for $r>1$ it is strong enough to localize all of the
electrons and create a gap in the spectral function.
Note also that 
as shown in the inset to Fig. 1, r is rapid function
of $\lambda$ in the region $r < 1$.  This rapid growth
of r means that the transition is very nearly first
order even for $n = 1/2$.
Figure 2 shows the temperature dependence of the resistivity
for the same coupling constants used to construct fig. 1.
The two lowest resistivities are calculated for weak couplings
$g < g_c$ and $g=g_c$,
for which $lim_{T \rightarrow 0} r^* (T) =0$
(so the $T=0$ spectral functions are identical).
For $g < g_c$, $\rho (T)$ at low $T$ is linear in $T$
with $g$-dependent prefactor and
$lim_{T \rightarrow 0} \rho (T)=0$.
For $g=g_c$ the quadratic term in the phonon energy
vanishes and $\rho \sim T^{1/2}$ at low $T$.
For $g$ slightly greater than $g_c$, $lim_{T \rightarrow 0}$
$\rho (T) \neq 0$ but is finite.
For $g$ large enough that the spectral function has a minimum
at $\omega =0$, the resistivity initially drops as $T$
is raised.
For $g$ large enough that the $T=0$ spectral function has a
gap, $\rho$ rises rapidly, and ultimately exponentially as
$T \rightarrow 0$.
The temperature at which $\rho$ begins to rise rapidly is
somewhat less than the $T=0$ gap.
Note that for $\sqrt{3\pi} /2 < g < 1.63$
$lim_{T \rightarrow 0} \rho (T) = \rho(0)$ is neither
zero nor infinite.

Some insight into the process of gap formation can be
gained from Figs. 3 and 4, which show the temperature
evolution of the spectral function and phonon
probability distribution for coupling $g= 1.69$.
One sees from fig. 3 that the rise in resistivity
as $T$ is decreased begins at the 
$T=0.07$ at which the spectral function
first develops a minimum.
A weak maximum in $P(x)$ is visible in fig. 4 even
at higher temperatures, but the maximum becomes an
obvious feature at the temperature at which the
resistivity begins to turn up.
The initial rapid rise in the resistivity is associated with
the development of the minimum in the spectral
function; the asymptotic low $T$ behavior $\rho \sim e^{\Delta /T}$ 
occurs at or below the lowest temperatures available numerically.
The numerical calculations in this section have all
been performed at $n=1/2$, where particle-hole
symmetry implies $\mu =0$, simplifying the computation.
Results for the particle-hole asymmetric case in other
models will be presented below.
In our computations to date we have found
that $n \neq 1/2$ has qualitatively similar behavior to
$n=1/2$ with metallic and insulating regimes according to
whether or not the $T=0$ spectral function has a gap.
As shown by the explicit calculations of Appendix C, somewhat
stronger couplings are required to obtain insulating
behavior at $n \neq 1$ than at $n=1$
In the numerical calculations at $T > 0$, we have seen
no evidence of the first order transitions found at $T=0$.
\subsection{Effect of Electron and Phonon Degeneracy}

\noindent
{\it Non-degenerate phonons}.
The effect of electron spin and orbital degeneracy on
models with scalar phonons is straightforward,
if there is no long range order.
One multiplies the quantity
Tr $ln [G_{eff}^{-1} ]$ in
Eq. \ref{eq:Z} by the factor
$n_{spin} n_{orb}$.
This factor may be absorbed into
$T$ and $g$
by defining $T=Tn_{spin} n_{orb}$, and
$g = g / \sqrt{n_{spin}n_{orb}}$ and the
model reduces to the previous case,
apart from small differences  coming from the $T$-dependence
of the fermion action.
This is shown in Fig. 6, which depicts $\rho (T)$
for a model with a two-fold electron degeneracy
($n_{spin} n_{orb}=2$).
The couplings are chosen to differ by a factor of $\sqrt{2}$ from
those used to construct Fig. 3, so as to produce the same
$T=0$ behavior.
The resistivity is smaller by a factor of two,
because it is proportional to $1/n_{orb} n_{spin}$.
If the temperature axis is rescaled by a factor of two,
the various curves lie almost on top of each other, with
agreement being best for low $T$ and weak coupling.
For example, the temperatures at which the resistivity turns up
for $g=1.83$ and 1.69 in the non degenerate case are
almost exactly one half of the temperatures at which $\rho$
turns up for $g=1.30$ and 1.20.
Some deviations from this scaling become apparent at higher
$T$; these are due to the $T$-dependence of the fermion action
(i.e. to the electron entropy).
For example, the scaled weak coupling resistivities differ by an
amount proportional to $T^2$.
Similarly, the upturn for the strongest coupling 
occurs at a slightly lower
rescaled temperature in the degenerate case.
Again for $g \gg g_c$ the model is insulating as $T \rightarrow 0$.
The approach to the $T \rightarrow 0$ limit 
for this model is shown in the
middle panel of Fig. 5.
It is seen that in this model our calculations
can access more of the low-T limit because of the rescaling
of the temperature.

\noindent
{\it Degenerate Phonons Jahn-Teller Coupled to Orbitally 
Degenerate Electrons}.
The theory is based on Eq. 5; significant differences from the
previous cases occur because the electron spectral function
may have a more complicated structure and the phonon measure
suppresses the probability of small amplitude lattice
distortions and increases the probability of large amplitude
distortions.

We begin with the effect of the phonon measure.
For this discussion we restrict attention to $n=1$,
so the only difference in the model is the factor of rdr in
the phonon measure.
Figure 7 shows the resistivity of this model for $n=1$ at
the same coupling constants used to construct Fig. 6.
One sees immediately that all resistivities are larger in the
Jahn-Teller case; also for parameters such that there is a
$T=0$ gap, up-turns occur at higher temperatures.
The origin of these differences may be seen most easily in
the phonon probability distribution $P(x)$, shown in Fig. 8 for
both degenerate and non degenerate phonons using in both cases
degenerate electrons.
We have chosen $g=g_c=1.085$ (the critical value at which the
$T=0$ lattice distortion vanishes) and $T=.15$.
The results are representative of all $g$ and $T$.
One sees immediately that the mean square value of
the lattice displacement is larger in the Jahn-Teller case
than in the non-degenerate case and more importantly the
small $r$ fluctuations are suppressed.
The larger mean square displacement means more
scattering and hence more resistivity.
The suppression of small $r$ fluctuations means that
$P(r)$ is reasonably sharply peaked at a non-zero value,
$r=r_{peak}$.
The temperature dependence of $r_{peak}$ in the
degenerate-phonon case is determined by the coupling;
for $g < g_c$ $r_{peak} (T) \sim T^{1/2}$, for
$g=g_c$ $r_{peak} (T) \sim T^{1/4}$ and for
$g > g_c$ $r_{peak}$ tends to a constant.

For non-degenerate phonons and $T=0$, $P(r)$ has a maximum at
$r=0$ if $g < g_c$ and maxima at $r \neq 0$ if
$g \gg g_c$.
Even for $g < g_c$ $P(r)$ at $T \neq 0$ may have a weak
maximum at $r \neq 0$, but this does not seem to have any
consequences for physical properties.
In the degenerate-phonon case the peak structure of $P(r)$
actually can lead to a minimum in the spectral function
at zero frequency if $gr_{peak}$ exceeds a critical value
somewhat larger than the 0.5 value found in Appendix C.
Of course if $g \lesssim g_c$, this minimum will vanish as T
is decreased.
This behavior is illustrated in Fig. 9, which shows the
temperature evolution of the spectral functions for the two
models.
Only the degenerate phonon case has a minimum at $\omega=0$;
the minimum vanishes below $T \sim 0.045$, 
at which temperature $gr_{peak} \sim 0.5$.
This may be understood by reference to the $T=0$
limit, in which $P(r)$ may be approximated by a delta function
at a coupling dependent value $r_0$.
If $gr_0 > 0.5$ the $T=0$ spectral function
develops a minimum at $\omega =0$; if $gr_0 > 1$ it develops
a gap.  Similarly, at $T>0$ the peak in the degenerate-phonon
$P(r)$ leads to a minimum in $A(\omega)$ if this peak occurs
at an $r_{peak}$ such that $gr_{peak} > 0.5$.
This tendency to open or increase a gap in the electron
spectral function as $T$ is raised acts to increase
the resistivity of the degenerate model above that of the
non-degenerate one.

Finally, the approach to the $T \rightarrow 0$ limit for the
insulating regime is shown in the upper panel of Fig. 5.
It is interesting that the result approaches the
$T \rightarrow 0$ limit more smoothly than in the two
previous cases, and also that $ T ln \rho$ is larger at
$T > 0$ than at $T=0$, unlike the other two cases.
We speculate that the origin of this difference is 
the larger rms value of $r$, which would imply an effectively
larger gap.

Away from $n=1$ a more important
difference in physics occurs, which is most easily seen for spinless
electrons in the very strong coupling limit.
Suppose $n < 1$ and neglect hopping completely.
Then by analogy with the steps leading to
Eq. \ref{eq:E_{gained}} one expects
n sites occupied by a single electron and 1-n
unoccupied.
The spectral function associated to the singly occupied
sites has two peaks, one below the chemical potential
corresponding to the occupied orbital and one
above corresponding to the unoccupied orbital.
The spectral function associated with the unoccupied
sites has only one peak, because in the absence of
any electrons there is no lattice distortion and
therefore no splitting of the orbitals.  Consider now the
first perturbative correction due to the hopping.
This will lead to a charge density $\delta n \sim n/\lambda$ on
previously unoccupied sites.  If (as assumed in our application
of the dynamical mean field method) there are no intersite
correlations in orbital occupancy or phonon fluctuations,
this extra charge will be randomly distributed over the
two on-site orbitals.  Thus in the presence of a
small amplitude phonon distortion $\delta r$ the charge density
will lead to an energy gain of at most $\delta n (\delta r)^2$,
which will be too weak to compete with the phonon stiffness
in the strong coupling limit. The three
peaked spectral function is therefore stable in
the strong coupling limit.  Appendices D and E derive
these results from an asymptotic analysis of the
dynamical mean field equations.
It is clear from the above arguments that the structure
of the mid-gap states may be affected by intersite correlations,
which could lead to a charge fluctuation favoring one
particular orbital, and therefore to an energy gain proportional
to $\delta r$ rather than $\delta r^2$.  Incorporating 
intersite correlations is an
important open problem. 

Of course, as $\lambda$ is decreased,
the possible energy gain increases and at some point the
mid-gap states split and the two pieces join the upper and lower
bands. The resulting intermediate coupling regime
corresponds to a uniform electron density and a uniform 
nonzero lattice 
distortion.  The electron spectral function consists of two bands,
with the fermi level in the lower one (for $n<1$).
The rigid band like state gains 
kinetic energy $\sim (1-n)$ relative to the three band state,
because one band is not fully filled, but is shown in
Appendices D and E to be more costly in Jahn-Teller energy
because the lattice distortion is not as large.

As the coupling is increased at $T=0$ three phases are in principle
possible:  a weak coupling phase with a single-peaked spectral
function, a "rigid band" phase with a two peaked
speactral function and the fermi level in the lower band, and
an insulating three peaked spectral function.
Determining the couplings at which the system
goes from one phase to another in general requires
numerical solution of the coupled mean field equations, 
along with comparison of the energies of different locally stable
solutions.  However, in the limit $(1-n) \rightarrow 0$
the equations simplify and a straightforward physical argument
may be used.  Consider the $n=1$ problem in the insulating regime
in which the spectral function has two peaks separated by a gap
and calculate the electron removal spectrum allowing the
on-site lattice distortion to adjust.
(The $\omega <0$ part of the spectral function previously
 computed for this
problem represents the lowest energy way to remove an
electron from the system at fixed lattice distortion.)
The energy may be computed from
the quantity $G^{ab}_{ii}(\omega,r)$ defined in
Eq. \ref{eq:bethe}, with the G  on the right hand side
of the equation given by the $n=1$ solution of the dynamical
mean field equations.  Performing this computation and optimizing over
r shows that there are two extrema--one at $r=0$ and one at $r=r^*$ 
(with $r^*$ the optimum $r$ found for $n=1$).  If $g>1.308...$
(i.e. $r^*>1.527...$ )
then the energy of the $r=0$ extremum is lower, showing that the
three peaked spectral function is favored, while for $g<1.308...$
the "rigid band" solution is favored. (Note that at
$g=1.308$ the two peaks of the rigid band spectral function
are spearated by a gap).  For this reason
we believe
that for $(1-n) \ll 1$ 
there is as $g$ is increased  a second
order transition to the rigid band two peaked
spectral function.  As $g$ is increased further, the
peaks separate and eventually a gap develops between
them (although the ground state remains conducting).
This is followed by a first order transition at a
larger $g$ to an insulating state with a three
peaked spectral function.
This sequence of spectral functions is shown in Fig. 10.
On the other hand,
for $n \rightarrow 0$ the results of Appendix E show that
the weak coupling one=peak spectral function proceeds directly
to the three peaked one via a first order transition.

One important consequence of the central feature
appearing in the three-peak regime is that
it acts to fill in the gap created by the lattice
distortion, so that one must go to stronger couplings and
lower temperatures to see insulating behavior in the
Jahn-Teller case than in the non-degenerate case.
This may be seen directly from the strong coupling
calculations of Appendices C and E.
In the non-degenerate case the spectral function
has two features, at $\pm g^2$ and physical quantities
are controlled by the energy gap between them, which is
of about the same size.
In the Jahn-Teller case the highest and lowest of the three
peaks are still  separated by an energy of $2g^2$, but
physical processes are controlled by the gap
separating the lowest and middle peaks, which is much less.
This may also be seen in numerical calculations of the low T
spectral functions and phonon probability distributions. The upper panel 
of Figure 11 compares the spectral functions for the
non-degenerate and Jahn-Teller cases at $T=0.03$, $g=1.58$ and
$n=0.75$.
Although the true $g > 1$ strong coupling limit has
not been reached, the two peak and three peaked structures 
expected for the non-degenerate and Jahn-Teller cases are
clearly visible and the gap in the Jahn-Teller case is much
less.  Similarly, the lower panel compares the spectral functions for
$n=1$ and $n-0.75$, for $g=1.58$, $T=0.03$ and Jahn-Teller coupling.
Again one sees the effect of the mid-gap states.
Figure 12 similarly compares the phonon probability 
distributions in the two cases.
Figure 13 compares the resistivities of the two models for the weaker
coupling $g=1.29$; the non-degenerate phonon model displays
insulating behavior below $T \sim 0.13$, while the
Jahn-Teller model remains metallic.
The resistivities cross at higher T because of the larger
fluctuations of $\langle r^2 \rangle$ in the Jahn-Teller case.
\section{Conclusions}
We have used the dynamical mean field method to study
the crossover from Fermi liquid to polaron behavior in
models of electrons coupled to localized classical
phonons.
The models we studied involved electrons with and
without spin and orbital degeneracy coupled to degenerate
and non-degenerate phonons.
We considered two forms of electron-phonon coupling:
conventional, in which the phonon displacement couples
to the electron charge density, and Jahn-Teller, in
which the phonon displacement couples to the splitting of the
electronic levels.
We showed that as $T \rightarrow 0$ three regimes
may be distinguished.
In {\it weak coupling} the mean square displacement of
the oscillator coordinate from its non-interacting
reference position vanishes and the momentum-integrated electron
spectral function assumes the non-interacting form.
Corrections lead to the usual linear $T$ resistivity.
In {\it intermediate coupling} the mean square displacement
of the oscillator coordinate is nonvanishing and the spectral
function is changed from its non-interacting form, but the
density of states at the Fermi level is nonzero so the
$T \rightarrow 0$ resistivity is neither zero nor infinite.
In {\it strong coupling} the mean square displacement is non-vanishing
and is sufficiently large that a gap appears in the spectral
function and the resistivity diverges as $T \rightarrow 0$.
Examples of the evolution of the $T=0$ momentum-integrated
spectral function with coupling strength are shown in
Figs. 1 and 10.
Figure 1 is calculated for the simplest model at
half-filling, but the results at any filling for
any model involving coupling of the oscillator to the
charge density of the electrons are similar.
The only important effects are that away from half-filling
in the strong coupling limit the relative weights of
the two peaks are different and that there is
at $T=0$ a discontinuous,
rather than continuous, change in the spectral function
and $\langle r^2 \rangle$ as the coupling is increased.
We believe, but have not proven, that at any
$T > 0$ the change is continuous.
The evolution of the spectral function for the case of
Jahn-Teller coupling is shown in Fig. 10.
The significant difference is the presence of the mid-gap
feature in the strong-coupling limit.
This has relative weight $| 2-2n|$ and corresponds physically
to sites unoccupied by electrons.
This mid-gap feature means that for Jahn-Teller couplings
and $n \neq 1$ the gap controlling physical properties
is much smaller at given coupling than it is in models
in which the phonon is coupled to the charge density,
and the mid-gap state is absent.

Three physical effects have been left out of our calculation.
One is the on-site Coulomb repulsion, which is certainly
important in transition metal oxides.
The Coulomb interaction tends to localize electrons, with the effect
being strongest at commensurate fillings and so will tend to
reinforce the localizing effect of the electron-phonon
interaction.  Because the localizing effect of the Coulomb
interaction is strongest at commensurate fillings, this will
also lead to an interesting doping dependence:  the
effective electron-phonon coupling will weaken as $n$
is varied from $1$.  The Coulomb interaction
will also change the form of the spectral function,
because it raises the energy of states with two electrons on
the same site.
For example, the uppermost peak of the strong-coupling
spectral function shown in Fig. 10 corresponds physically to
the density of states for adding an electron to a site
with one electron already present, at fixed phonon
configuration.
Coulomb effects will shift this energy upwards.
A quantitative treatment of these effects within the present
formalism will be left to a future paper.

A second important piece of physics is quantum fluctuations of the
phonons.
At low temperatures and in the absence of long range order,
these will allow electrons to move from site to
site even in the strong coupling regime, so cutting off the
strong-coupling divergence in the resistivity.
One mathematical consequence is the appearance of Gaussian
tails to the spectral functions in the gap regions.

A third omitted piece of physics is long range order.
It has been assumed throughout that the lattice
distortions are random from site to site; thus
the localization is due purely to polaronic effects.
Near commensurate densities, charge density wave
effects will be important as well.

\noindent
{\it Acknowledgements}:
We thank P. B. Littlewood, B. G. Kotliar and A. Sengupta
for helpful discussions.
R. M. was supported in part by the Studienstiftung
des Deutschen Volkes.
\begin{appendix}
\section{Relation to Migdal-Eliashberg Theory}

In this Appendix we derive the relationship between the
formalism we have used and the conventional Migdal-Eliashberg theory
of the weakly coupled electron-phonon system.
The essential parameter of the conventional treatment is a
dimensionless coupling constant $\lambda_{conv}$ which is defined
in terms of the electron self energy $\Sigma (\omega ,T)$ via
\begin{equation}
lim_{\omega \rightarrow 0} \frac{\partial \Sigma (\omega ,T=0)}
{\partial \omega} =\lambda_{conv} .
\label{eq:lim}
\end{equation}

To establish the relationship we introduce a finite phonon mass
$M_{ph}$.
This implies a Debye frequency $\omega_D$ given by
\begin{equation}
\omega_D^2 = k /M_{ph} a^2
\label{eq:omega_D}
\end{equation}
with $a$ the lattice constant.
(We use units in which $\hbar =1$.)
We then introduce phonon creation and
annihilation operators
$b^{\dagger} ,b$ via
\begin{equation}
r = (b^{\dagger} + b) /(2M_{ph} \omega_D)^{1/2}
\label{eq:x}
\end{equation}
Comparison to Eq. \ref{eq:H_{el-ph}^{charge}} yields
\begin{equation}
H_{el-ph}^{conv} = \left ( \frac{g^2 }{2M_{ph}\omega_D} \right )^{1/2}
\sum_{ia\alpha} (d_{ia\alpha}^+ d_{ia\alpha} -n)
(b_i^+ + b_i )
\label{eq:H^{conv}}
\end{equation}

The standard electron-phonon calculation gives
\begin{equation}
\frac{\partial \Sigma (\omega , T=0)}{\partial \omega} =
n_{orb} n_{spin} \frac{g^2}{k} \cal 
D(\epsilon_F)
\label{eq:partialSigma/partialomega}
\end{equation}
where $\cal D(\epsilon_F)$ is the single-spin density of states
at the Fermi surface.
Thus
\begin{equation}
\lambda_{conv} = t \cal D(\epsilon_F) \lambda
\label{eq:lambda_{conv}}
\end{equation}

Elsewhere in this paper it has been shown that in the dynamical
mean field calculations polaron effects occur at $\lambda \sim 1$
implying $\lambda_{conv} \sim 1$.
For example, in the model with $n_{orb}=n_{spin}=1$ and a
semicircular density of states at $\mu =0$, frozen phonon
distortions begin to occur at $\lambda = \pi$ ($\lambda_{conv}=1)$.
The critical values of $\lambda$ found here are articifically
small because of neglect of quantum fluctuations and
intersite interactions.
In models with quantum phonons, somewhat larger $\lambda$ values
will be required to produce insulating behavior, but the general
conclusion that metallic behavior breaks down at a
$\lambda_{conv}$ not too much larger than $1$ will still hold.
Now Migdal showed, using phase-space arguments,
that the parameter which controls perturbation theory
about the zero electron-phonon coupling limit is not $\lambda$ but
$\lambda \omega_D /t$ (or if $T > \omega_D$,
$\lambda T /t)$ \cite{Migdal58}.
Therefore, one might expect to be able to use the Migdal-Eliashberg
equation to study the crossover from Fermi liquid to polaron physics.
The difficulty with this argument, however, is that
the ground state about which to perform the expansion
is not known for $\lambda > 1$.
In fact, in the limit considered in this paper, the Migdal-Eliashberg
equations are identical to those obtained by expanding
Eq. \ref{eq:Z} to order $r^2$ and solving self-consistently.
Corrections to this self-consistent solution are indeed small,
if $\lambda T/t$ is small, but the starting point is seen to 
be wrong if
$\lambda > \lambda_c \approx 1$.

\section{Non-degenerate Fermions
with Lorentzian Density of States}

In this Appendix we
use steepest descents techniques to analyze the $T \rightarrow 0$
limit of spinless fermions with a Lorentzian density of
states coupled to a scalar classical oscillator.
A similar technique was used in Ref \cite{Freericks93}.
The partition function is
\begin{equation}
Z( g,n) = \int dr e^{-E(r,g,n)/T}
\end{equation}
with
\begin{equation}
E(r,g,n) = \frac{1}{2} r^2 -
g rn- [\mu-gr ][ \frac{1}{2} + \frac{1}{\pi}
tan^{-1} ( \mu - g r)]
+ \frac{1}{2 \pi} ln [1+ ( \mu - g r)^2 ]+ \mu n
+ {\cal O} T^2.
\label{eq:Elor}
\end{equation}
and the chemical potential $\mu$ chosen so that
\begin{equation}
dZ/d \mu = 0.
\label{eq:dZ}
\end{equation}

As $T \rightarrow 0$, the r integral is dominated
by the values $r^*$ minimizing E.
We have
\begin{equation}
Z (g,n) \rightarrow \sum_a Z_a exp-E(r_a^*,g,n) /T.
\label{eq:Zlor2}
\end{equation}
Here a labels the extrema; $r_a^*$ is a solution of
\begin{equation}
r=g(n-\frac{1}{2} - \frac{1}{\pi} tan^{-1}
( \mu - gr))
\label{eq:rlor}
\end{equation}
and $Z_a \geq 0$ is the weight associated with extremum $a$.
Each $Z_a$ is a product of two contributions, one from integrating over
the quadratic fluctuations in $r$ about $r=r_a^*$ and one from the leading
$T$-dependence of $\mu$.
In the present problem we have, at low $T$, $\mu = \mu_0+AT$;
the $T$-linear term contributes to $Z_a$.

Equation \ref{eq:rlor} may have one or three solutions.
In the latter case, either one has the lowest energy or two are degenerate.
If there is one dominant extremum then Eq. \ref{eq:dZ} implies
\begin{equation}
\mu = gr+ tan \pi (n-1/2).
\label{eq:mu1}
\end{equation}
so Eq. \ref{eq:rlor} implies $r=0$.

Now consider the case of two degenerate extrema, at $r=r_1$ and
$r-r_2$.  From Eq. \ref{eq:dZ} and Eq. \ref{eq:rlor} one finds
$Z_1r_1 + Z_2r_2=0$; as $Z_{1,2} \geq 0$, $r_1$ and $r_2$ must have
opposite signs.
It is convenient to define
\begin{equation}
\begin{array}{rl}
R &= g (r_1+r_2)/2, \\
\Delta &= g(r_1-r_2)/2.
\label{eq:R}
\end{array}
\end{equation}
In terms of these variables Eqs. \ref{eq:rlor}, become
\begin{equation}
R = g^2(n-1/2) - \frac{g^2}{2\pi} tan^{-1}
\frac{2\mu -2R}{1-(\mu - R)^2+ \Delta^2}, \\
\label{eq:Rlor2}
\end{equation}
\begin{equation}
\Delta = \frac{g^2}{2\pi} tan^{-1} 
\frac{2\Delta}{1+(\mu - R)^2-\Delta^2}.
\label{eq:dellor2}
\end{equation}

These equations must be solved subject to the constraint
$E(r_1,g,n)=E(r_2,g,n)$ which by use of Eqs. \ref{eq:Rlor2},
\ref{eq:Elor} may be written
\begin{equation}
(R- \mu) \Delta =
\frac{g^2}{4\pi} ln \frac{1+(R- \mu + \Delta )^2}
{1+(R- \mu - \Delta )^2}.
\label{eq:consistlor}
\end{equation}
At small $R$ and $\Delta$, 
Eqs. \ref{eq:consistlor} and \ref{eq:dellor2} are only consistent
if $R=\mu = g^2(n-1/2)$, so \ref{eq:consistlor} is trivial.
This may be most easily seen by expanding the two equations in
$\Delta$ at fixed $\mu -R$ and comparing the $\Delta^3$ coefficients.
This immediately implies that if $n \neq 1/2$ the transition is first
order.

In the strong coupling limit, $r_1=-g(1-n)$,
$r_2=gn$ and $\mu=\frac{1}{2}g^2 (n-1/2)$.
\section{Non-degenerate Fermions with Semicircular
Density of States}

In this Appendix we analyze the $T \rightarrow 0$
limit of non-degenerate fermions with a semicircular
density of states coupled to a classical oscillator.
The technical differences from the Lorentzian case treated in
Appendix B are that the mean field parameter satisfies the
self-consistency equation Eq. \ref{eq:a_{semi}} and that the
energy is given by
\begin{equation}
\begin{array}{rr}
E(r,g,n) = - \frac{1}{2} T \sum_n [a_n-(i\omega_n+\mu )]^2 
+ \frac{r^2} {2} - grn - T \sum_n ln [a_n - gr] + \mu n.
\label{eq:Esemi}
\end{array}
\end{equation}
with $r$ given by a solution of
\begin{equation}
r=g(n+ \sum_n \frac{1}{a_n+gr}).
\label{eq:rsemi}
\end{equation}
If there is one dominant extremum then Eqs. \ref{eq:dZ} and
\ref{eq:a_{semi}} imply $n=- \sum_n(a_n+gr)^{-1}$
so Eq. \ref{eq:rsemi} implies $r=0$.
In this case
\begin{equation}
a(i\omega_n) = \frac{1}{2}
[(i \omega_n+ \mu ) -sgn \omega_n
\sqrt{(i\omega_n+\mu )^2 -4} ].
\label{eq:a0}
\end{equation}

If there are two degenerate extrema, Eqs. \ref{eq:Zlor2},
\ref{eq:R} apply, as do the relations
$Z_1r_1 + Z_2r_2=0$ and $E(r_1,g,n)=E(r_2,g,n)$ used in
Appendix B.
It is convenient to rewrite the equations in terms of
$R,\Delta$, $b_n=a_n+R$ and $z_n=i\omega_n+ \mu +R$ as
\begin{equation}
R = g^2n + g^{2} T \sum_n \frac{b}{(b^2-\Delta^2)}, \\
\label{eq:nsemi}
\end{equation}
\begin{equation}
\Delta = -g^2 T \sum_n \frac{\Delta}{(b^2-\Delta^2)}, \\
\label{eq:delsemi}
\end{equation}
\begin{equation}
2R\Delta = 2g^2 n\Delta + g^2 T \sum_n ln
\frac{b+\Delta}{b-\Delta}, \\
\label{eq:consistsemi}
\end{equation}
\begin{equation}
b = z - \frac{b+R}{b^2-\Delta^2}.
\label{eq:b}
\end{equation}

For $n \neq 1/2$, the transition may be shown to be first order
by the argument used in Appendix B:
one expands Eqs. \ref{eq:delsemi} and \ref{eq:consistsemi} to
order $ \Delta^3$ at fixed $b$ and $R$ and observes that the
equations are not compatible.

For $n=1/2$, $\mu =R$ and $b$ is odd in $i\omega_n$, 
so Eqs. \ref{eq:nsemi},
\ref{eq:consistsemi} are satisfied trivially.
The other two equations may be easily solved by viewing them as
equations for $b$ and $g^2$ at given $\Delta$.
Because Eq. \ref{eq:b} is only cubic, the solution may be written down
immediately.
It is easist to find the proper branch of the solution if $Z$
is continued to the real axis.
Once $b$ is found, $g^2$ may easily be computed 
from Eq. \ref{eq:delsemi}.  From Eq. \ref{eq:b} 
one sees immediately that the imaginary part of
$b$, $b^{''}$, vanishes at $\omega=0$ for $\Delta^2=1$.
The coupling corresponding to this, $g= 1.63$ is the one at which
a gap first appears in the spectrum.
Similarly, one may show that the leading ($\omega^2$) correction to the
$\omega=0$ value of $b^{''}$ vanishes at $\Delta^2=1/4$.

The strong coupling limit is analytically tractable at all $n$.
$b^{''}$ is nonzero only for $Z$ near $\pm \Delta$.
Defining $b_{\pm} = b \pm \Delta$ and
$Z_{\pm}=Z \pm \Delta$ and neglecting terms of order
$1/ \Delta$ one finds from Eq. \ref{eq:b}
\begin{equation}
b_{\pm} \cong Z_{\pm} - \frac{1}{2b_{\pm}} (1 \pm R/ \Delta).
\label{eq:b+-}
\end{equation}
This equation may be solved and the results inserted into the
other equations.
One finds
\begin{equation}
\begin{array}{rl}
r_1 &= -g(1-n), \\
r_2 &= gn, \\
\mu &= g^2 (n-1/2).
\label{eq:rstrong}
\end{array}
\end{equation}

The physical content of these solutions is one band, of width $\sqrt{2 n}$
centered at $\omega =- g^2$ representing occupied states and another
at $\omega =g^2$ of width $\sqrt{2(1-n)}$ representing unoccupied states.
\section{Orbitally Degenerate Electrons with
Lorentzian Density of States}

In this Appendix we analyze the $T \rightarrow 0$ limit of
orbitally degenerate electrons with a Lorentzian density of
states coupled to a classical Jahn-Teller oscillator whose
displacement has magnitude r.
The treatment parallels Appendix B but the details are different.
Instead of B-2 we have
\begin{eqnarray*}
E(r,g,n)= \frac{1}{2} r^2-
(\mu +gr)(\frac{1}{2} +\frac{1}{\pi} tan^{-1} (\mu+gr)) \\
-(\mu-gr)(\frac{1}{2}+\frac{1}{\pi} tan^{-1}(\mu-gr)) +
\frac{1}{2 \pi} ln [1+(\mu+gr)^2] \\
+\frac{1}{2\pi} ln [1+(\mu- gr)^2]+\mu n
\label{eq:Elordegen}
\end{eqnarray*}
with $\mu$ chosen so B-3 holds.
Instead of Eq. \ref{eq:rlor} we have
\begin{equation}
r= \frac{g}{\pi} [tan^{-1} ( \mu +gr) - tan^{-1} (\mu -gr) ].
\label{eq:rdegen}
\end{equation}

Suppose first there is only one dominant extremum.
If $g$ is small this is at $r=0$ and
\begin{equation}
\mu=tan \frac{\pi}{2} (n-1)
\end{equation}
and
\begin{equation}
E(r=0,n)=- \frac{2}{\pi} ln
(cos \frac{\pi}{2} (n-1)).
\end{equation}
At $g=g_c(\mu)= 2/ (\pi(1+\mu^2)$, there is a second order transition
to a state with $r \neq 0$.
In this state, the chemical potential is fixed by
\begin{equation}
n=1+ \frac{1}{\pi}
[tan^{-1} (\mu +gr) +tan^{-1}(\mu-gr)].
\label{eq:ndegen}
\end{equation}
As $g \rightarrow \infty$ at $n \leq 1$, this solution tends 
to $r=\pm gn$, $\mu= tan \pi (n-1/2) -g^2 n$,
$E=- \frac{1}{2} g^2n^2 + \frac{1}{\pi} ln g^2n$.

At large $g$ and $n \neq 1$, Eq. \ref{eq:rdegen} has an 
alternative solution, $r=g$.
This is incompatible with Eq. \ref{eq:ndegen}, so the solution must be
degenerate with another extremum which, by symmetry, must be at $r=0$.
These requirements imply that $\mu = \frac{1}{2} g^2 sgn (n-1)$ and
$E \rightarrow - \frac{1}{2} g^2 n+ \frac{4}{\pi} ln g$.
If $n$ is near 1 the two extremum solution becomes lower
in energy at $g \sim 1/ \sqrt{1-n} \gg g_c$; if $n$ is near 0,
the two extremum solution is lower in energy at all $g >g_c$.

\section{Jahn-Teller Coupled Electrons with Semicircular Density
of States}

In this Appendix we analyze the $T \rightarrow 0$ limit
of orbitally degenerate electrons with a semicircular density
of states coupled to a classical Jahn-Teller oscillator whose
displacement has magnitude r.
The treatment follows Appendix C but with technical differences
analogous to those found in Appendix D.
Instead of Eq. \ref{eq:Esemi} we have
\begin{equation}
E(r,g,n)=T \sum_n
[a_n-(i\omega_n+\mu)]^2 
+ \frac{1}{2} r^2
- T \sum_n ln [a_n^2- g^2r^2] +\mu n.
\end{equation}
Instead of Eq. \ref{eq:rsemi} we have
\begin{equation}
r=-2g^2rT \sum_n \frac{1}{a_n^2-g^2r^2}.
\label{eq:rdegensemi}
\end{equation}
If there is one dominant extremum then $a$ satisfies
\begin{equation}
a_n=i \omega_n+ \mu - \frac{a_n}{a_n^2-g^2r^2}
\label{eq:adegen1}
\end{equation}
while $\mu$ is given by
\begin{equation}
n=2T \sum_n \frac{a_n}{a_n^2-g^2r^2}.
\label{eq:ndegen1}
\end{equation}

For small $g$, the only possible solution is $r=0$, so $a_n$ is given by
Eq. \ref{eq:a_{semi}} and $\mu$ may be found numerically from
Eq. \ref{eq:ndegen1}.
The $r=0$ solution becomes linearly unstable at
\begin{equation}
g_c^2 ( \mu )= 3 \pi /(4-\mu^2)^{3/2}
\label{eq:gcdegen}
\end{equation}
to a solution with $r \neq 0$.
This solution implies a frozen in Jahn-Teller splitting random
from site to site but of the same magnitude on all sites,
and therefore a two peaked spectral function.
Again the equations simplify in the strong coupling limit, where the
reasoning that led to Eq. \ref{eq:b+-} shows that the single-extremum
solution tends to one with a spectral function consisting of two
semicircular features of half width $\sqrt{2}$ (vs half-width 2 at
$g <g_c$), centered at $\omega= - \mu \pm g^2 n$.
For $n <1$ $\mu$ is such that the lower band is partly
filled, thus $\mu =- g^2n+ {\cal O} (1)$, and the energy is
$- \frac{1}{2} g^2n^2 +  {\cal O} (1-n)$.
the ${\cal O} (1-n)$ term comes from carrier kinetic energy 
in the partly
filled band.

As in Appendix D, an alternate two-extremum solution exists, with
one extremum at $r=0$ and the other at $r \neq 0$.
As $g \rightarrow \infty$, the nonzero $r \rightarrow g$.
The spectral function consists of three well separated
peaks; two each of weight $min (n, 2-n)$, at
$\omega -\mu = \pm g$ corresponding to sites with a electron and one, of
weight $2 |1-n|$, centered at $\omega =-\mu$ and corresponding to the
empty sites if $n <1$ and the doubly occupied sites at $n >1$.
The chemical potential is $- \frac{1}{2} g^2$ and the energy is
$- \frac{1}{2} g^2 n+ {\cal O} 1/g$.  As $g \rightarrow \infty$ the two
extremum solution is lower in energy.  For $n \rightarrow 0$ the divergence
of $g_c$ shown in Eq \ref{eq:gcdegen} implies that the undistorted
state proceeds directly to the three peaked 
one via a first order transition.   For $n$ near 1 the calculation
sketched in the text shows that the two-peaked solution, 
with a gap between the two
peaks, exists for a finite range of $g$
\end{appendix}

\newpage
\section*{Figure Captions}
\begin{itemize}
\item[Fig. 1]
Momentum-integrated electron spectral function $A(\omega )$
plotted against frequency $\omega$ for non-degenerate electrons
coupled to non-degenerate phonons at density $n=1/2$,
temperature $T=0$ and several different values of frozen-in
lattice distortions $r=0$ (heavy dashed line), .87
(heavy dotted line), 1.0 (light solid line), 1.17 (light dashed line)
and 1.49 (light dotted line)
corresponding to $g=1.53$, 1.60, 1.63, 1.69 and 1.83, respectively.
Inset: coupling constant ($\lambda$) dependence of $r$.
\item[Fig. 2]
Temperature ($T$) dependence of resistivity $\rho$ calculated
for non-degenerate electrons coupled to non-degenerate phonons at
density $n=1/2$ for  the
coupling constants used to create Fig 1:  $g=1.29$ (heavy solid line),
$1.53$ (heavy dashed line), $1.60$ (heavy dotted line),
$1.63$ (light solid line),
$1.69$ (light dashed line),   and $1.83$ (light dotted line).
Note that $g=1.29$ and $g=1.53$ have the same lattice distortion $r=0$
at T=0 and therefore the same spectral function.
\item[Fig. 3]
Temperature ($T$) dependence of electron spectral function
$A(\omega )$ for $g=1.69$, $n=1/2$ and
$T=0$ (light solid line), .01 light dashed line), .02 (light dotted line),
.07 (heavy solid) and .15 (heavy dashed line).  Inset:
spectral function plotted over a wider range of frequency for
parameters used in main figure.  The higher the T the longer the
high frequency tail.
Comparison to the appropriate curve in Fig. 2 shows
that the rise in the resistivity begins at the $T =0.07$ at which 
$A$ first begins to develop a minimum at $\omega = 0$.
\item[Fig. 4]
Temperature dependence of phonon probability distribution
$P(r)$ for $g=1.69$,
$n=1/2$, and $T=.01$ (light dashed line), .02 (light dotted line),
.07 (heavy solid line) and .15 (heavy dashed line).
\item[Fig. 5]
Low temperature behavior of resistivity for the three
models considered in the text for parameters such that the 
spectral function has a gap at $T=0$:
lower panel: non-degenerate phonons and non-degenerate electrons, 
middle panel: non-degenerate
phonons and degenerate electrons, upper panel: degenerate phonons
and degenerate electrons.
The expected low-$T$ behavior is $\rho \sim T^x e^{\Delta /T}$
with $\Delta$ the low-$T$ gap in the spectral function.
The heavy dots mark values of $\Delta$ obtained from an analytic $T=0$ calculation
as described in the
text.
\item[Fig. 6]
Temperature ($T$) dependence of resistivity
($\rho$) for model of degenerate electrons and non-degenerate
phonons with $n=1$
and $g=0.91$ (heavy solid line), $1.09$ (heavy dashed line),
$1.13$ (heavy dotted line), $1.15$ (light solid line),
$1.20$ (light dashed line) and $1.30$ (light dotted line).
These couplings produce the same $T=0$ gaps as in Fig. 2.
The factor of $\sqrt{2}$ in the couplings and the factor of 2 in the 
resistivity relative to Fig. 2 comes from orbital degeneracy.
\item[Fig. 7]
Temperature ($T$) dependence of resistivity ($\rho$) for model
of degenerate electrons and degenerate phonons with $n=1$
and $g=0.91$ (heavy solid line), $1.09$ (heavy dashed line),
$1.13$ (heavy dotted line), $1.15$ (light solid line),
$g=1.20$ (light dashed line) and $1.30$ (light dotted line).
These couplings produce the same $T=0$ gaps as in Figs. 2, 6.
Note $\rho$ is much larger than that shown in fig 6
for the non-degenerate
electrons model, reflecting the stronger scattering.
\item[Fig. 8]
Lower panel:
phonon probability distribution $P(r)$ for $g=1.085$,
$n=1$, $T=.05$ (heavy solid line), .1 (light solid line) and 
.15 (light dashed line)
and non-degenerate phonons.
Upper panel:
phonon probability distribution $P(r)$ for same
parameters and degenerate phonons.
\item[Fig. 9]
Comparison of temperature evolution of electron spectral
functions for degenerate (upper panel) and non-degenerate
(lower panel) phonons and orbitally degenerate electrons
for $n=1$ and $g=1.085$, $T=.05$ (heavy solid line), 
.1 (light solid line) and .15 (light dashed line) as used in Fig. 8.
\item[Fig. 10]
Evolution of $T=0$ spectral function with coupling
strength. Model with degenerate electrons, degenerate phonons and
$n=0.75$ for coupling constants $g =1.53$ (light solid line), 
1.58 (light dashed line), and the strong coupling limit solution evaluated
at $g = 2.47$ (heavy solid line).
\item[Fig. 11]
Comparison of low $T$ spectral functions. Upper panel Jahn-Teller
(heavy line, shifted by $\mu = -1.23$) and non-degenerate phonons
(light line) at $T=.03$ $g=1.58$, and $n=0.75$.
Lower panel: $n=1$ (light) and $n=0.75$ (heavy) with Jahn-Teller phonons,
$T=0.03$ and $g=1.58$.
$n=0.75$ curve (heavy line) shifted by $\mu =-1.23$.
\item[Fig. 12]
Comparison of low $T$ phonon probability distribution
$P(r)$ for Jahn-Teller (heavy line) and non-degenerate
(light line) phonons for $T=.03$, $g=1$, and $n=0.75$.
Note that the large $r$ peaks occur at almost exactly the same r.
Note also that the Jahn-Teller $P(r)$ is defined only for
$r> 0$.
The asymmetry in $P(r)$ in the non-degenerate case is a
consequence of particle hole assymmetry due to $n \neq 1$.
\item[Fig. 13]
Comparison of temperature dependent resistivities
for Jahn-Teller (heavy line) and non-degenerate phonons
(light line), $g=1.29$ and $n=0.75$.
\end{itemize}
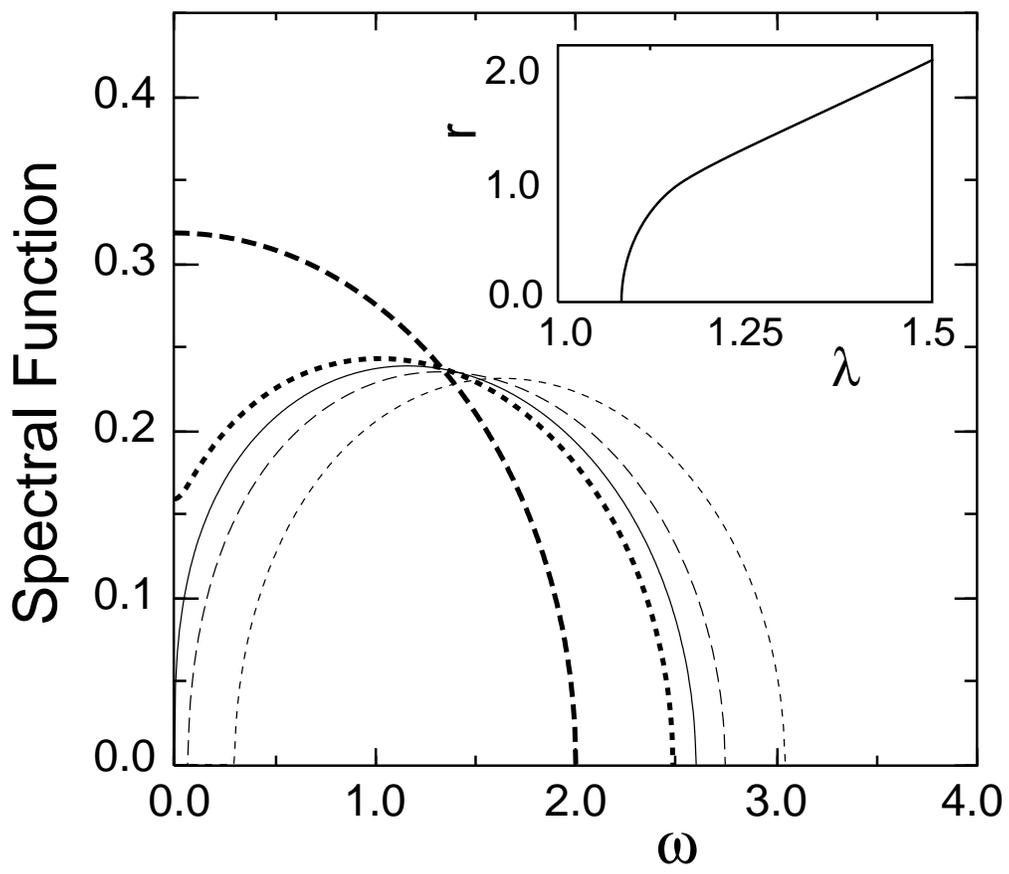
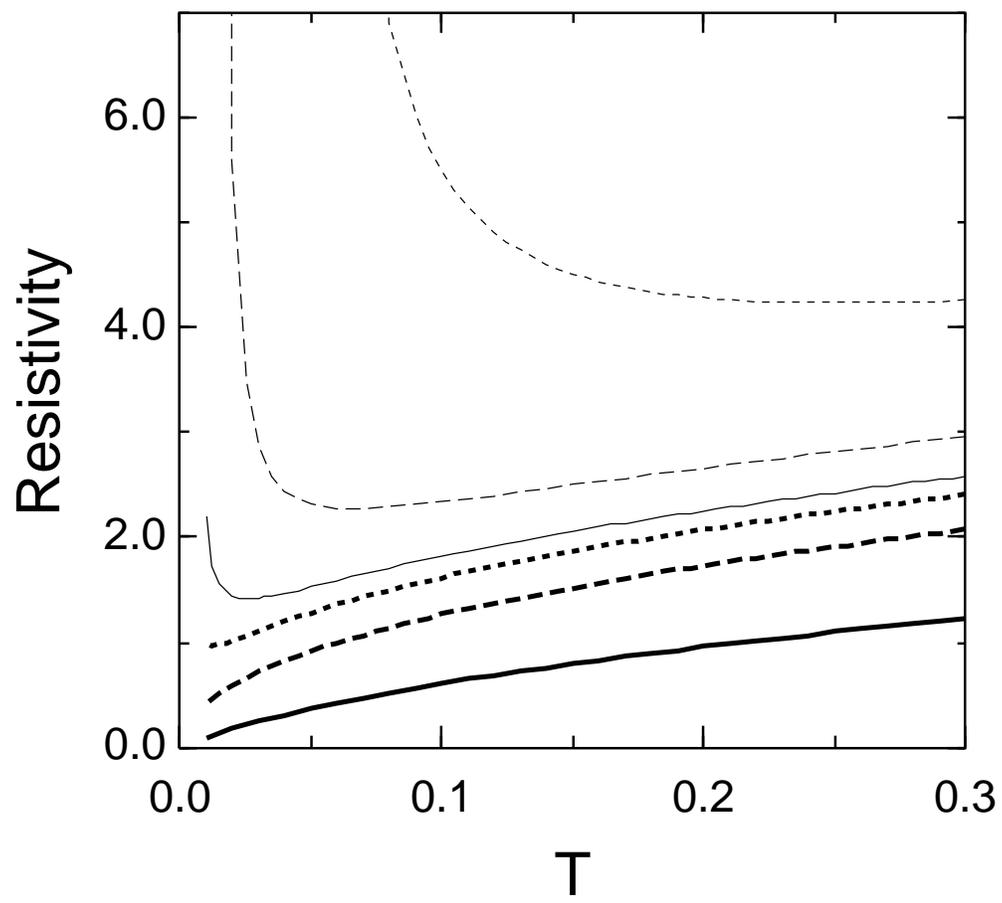
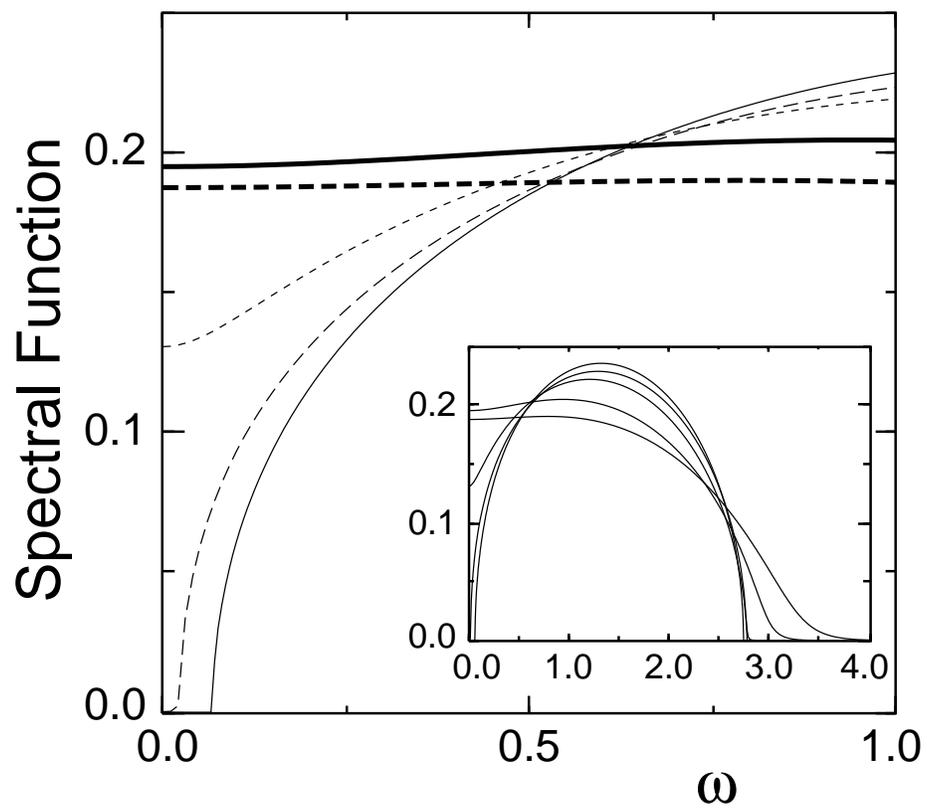
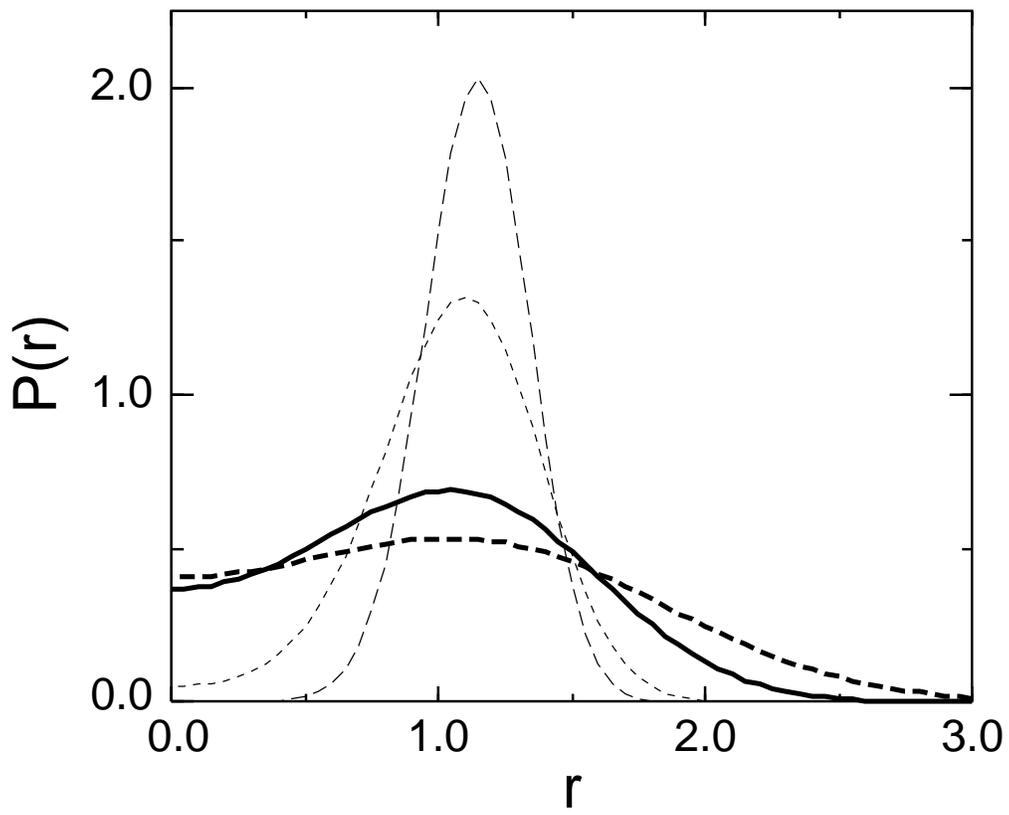
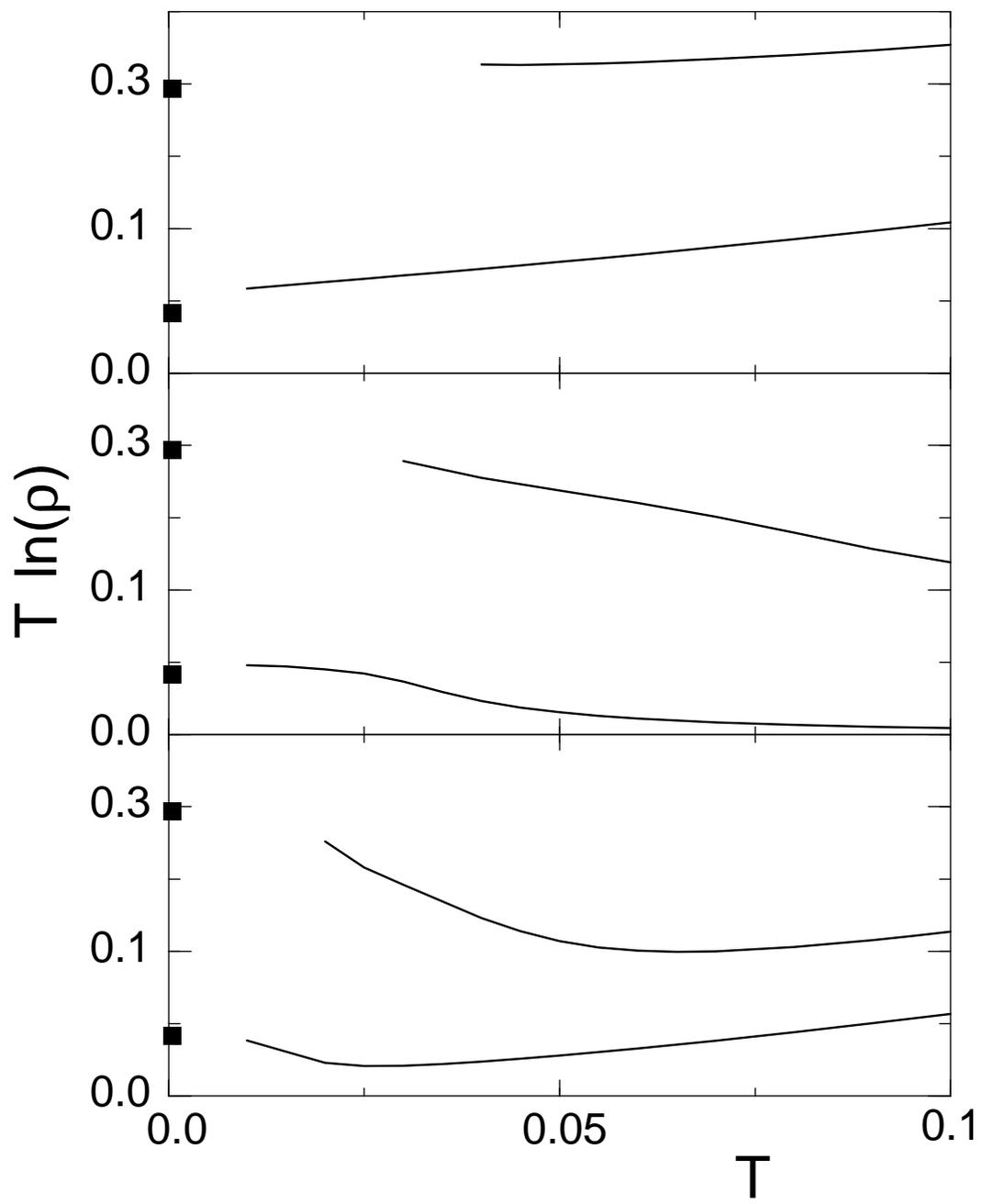
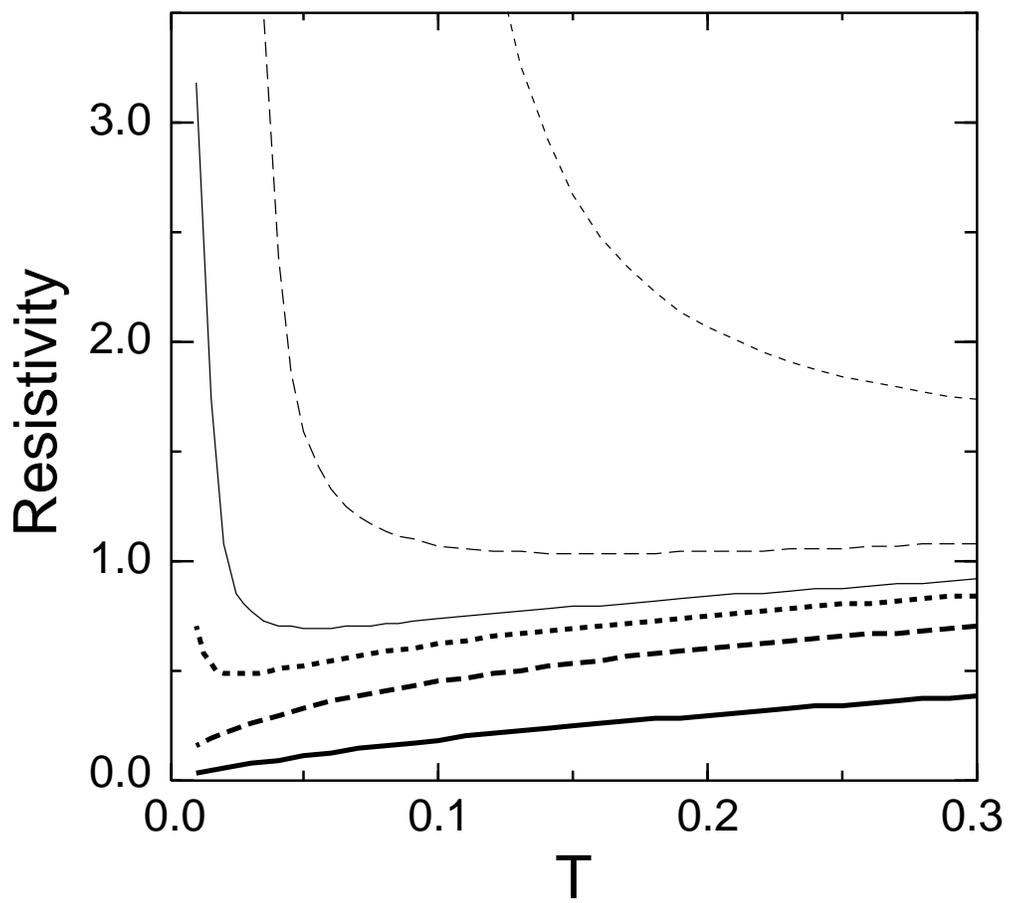
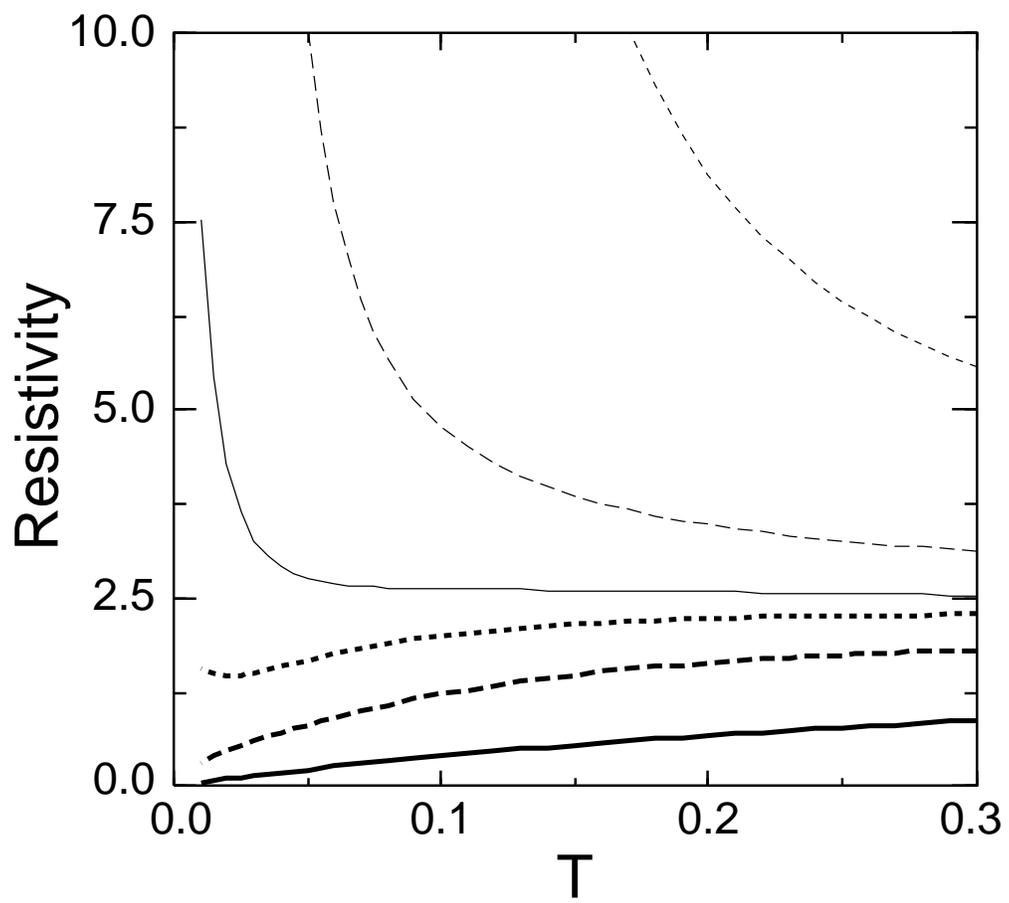
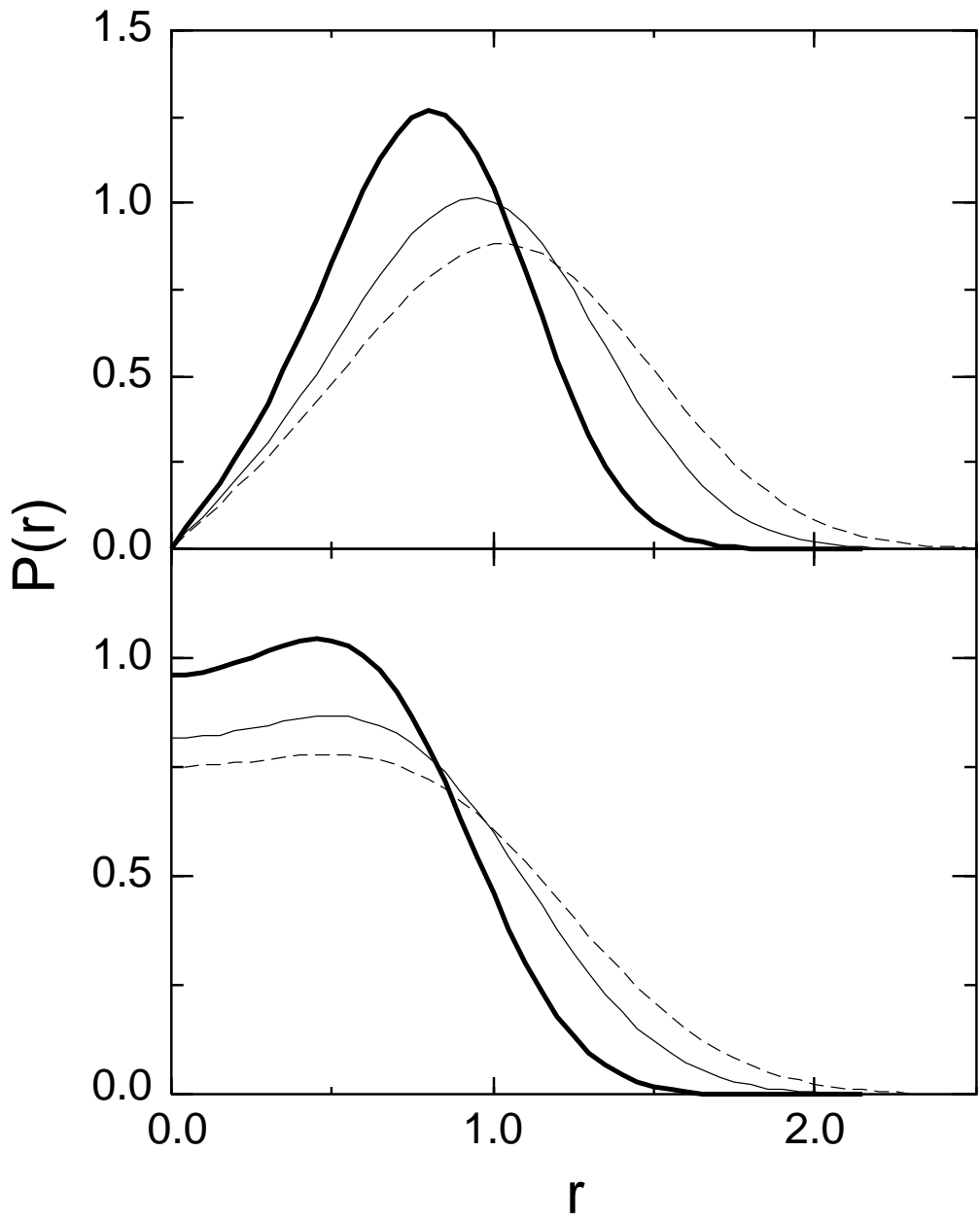
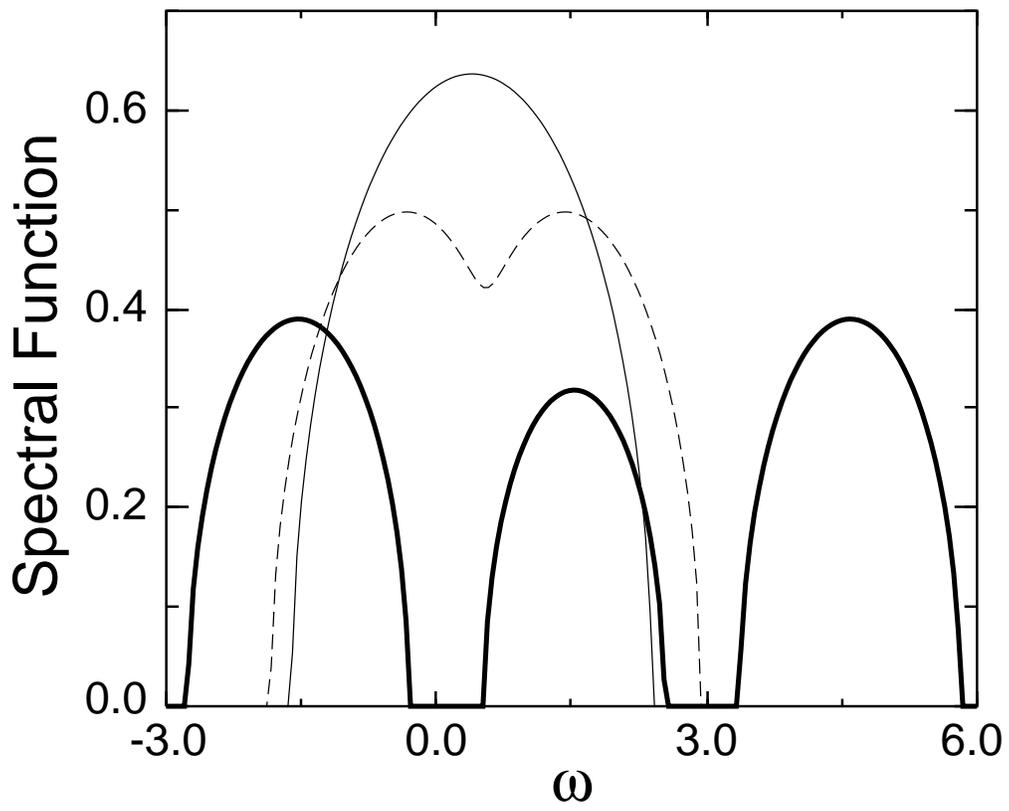
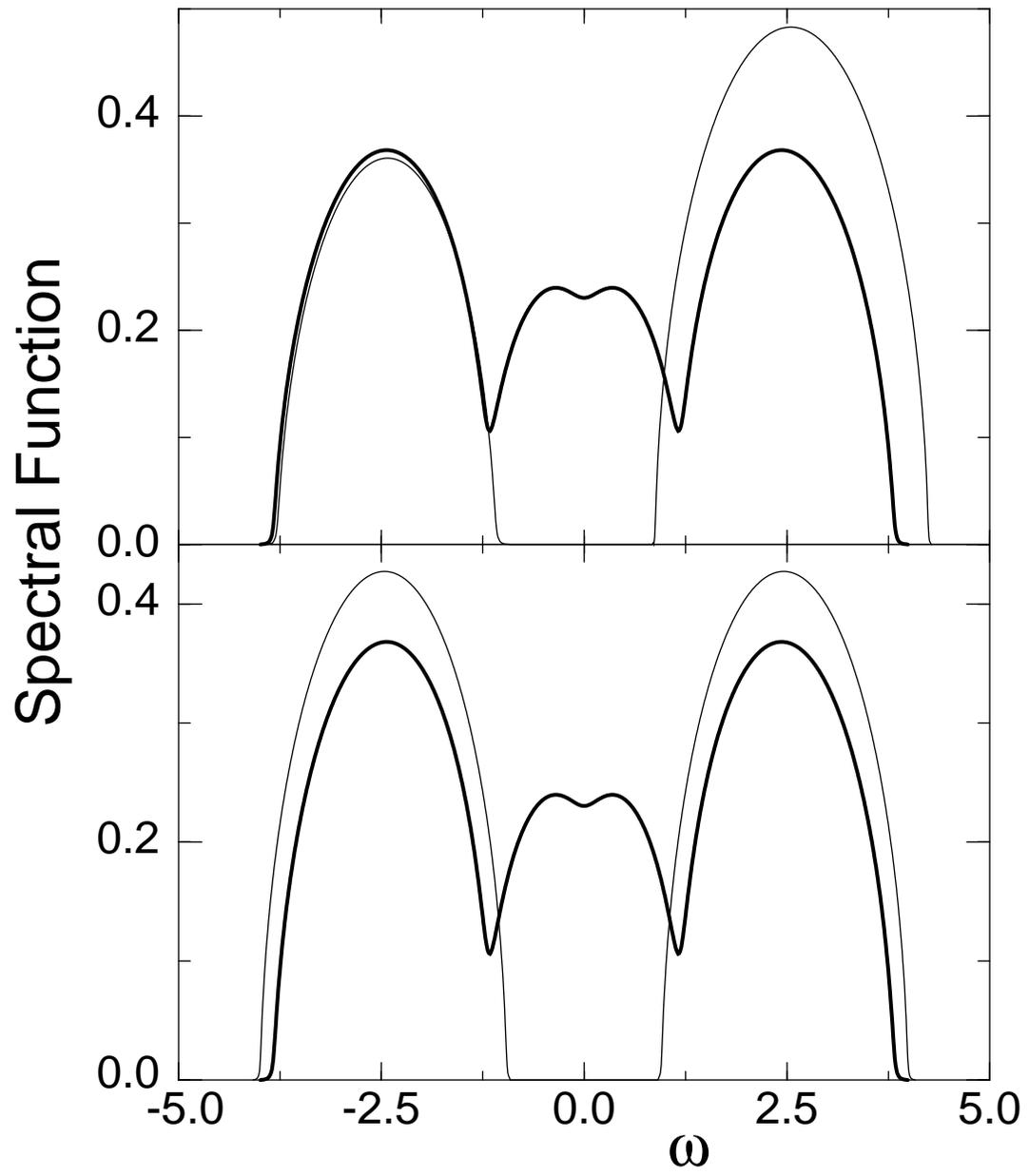
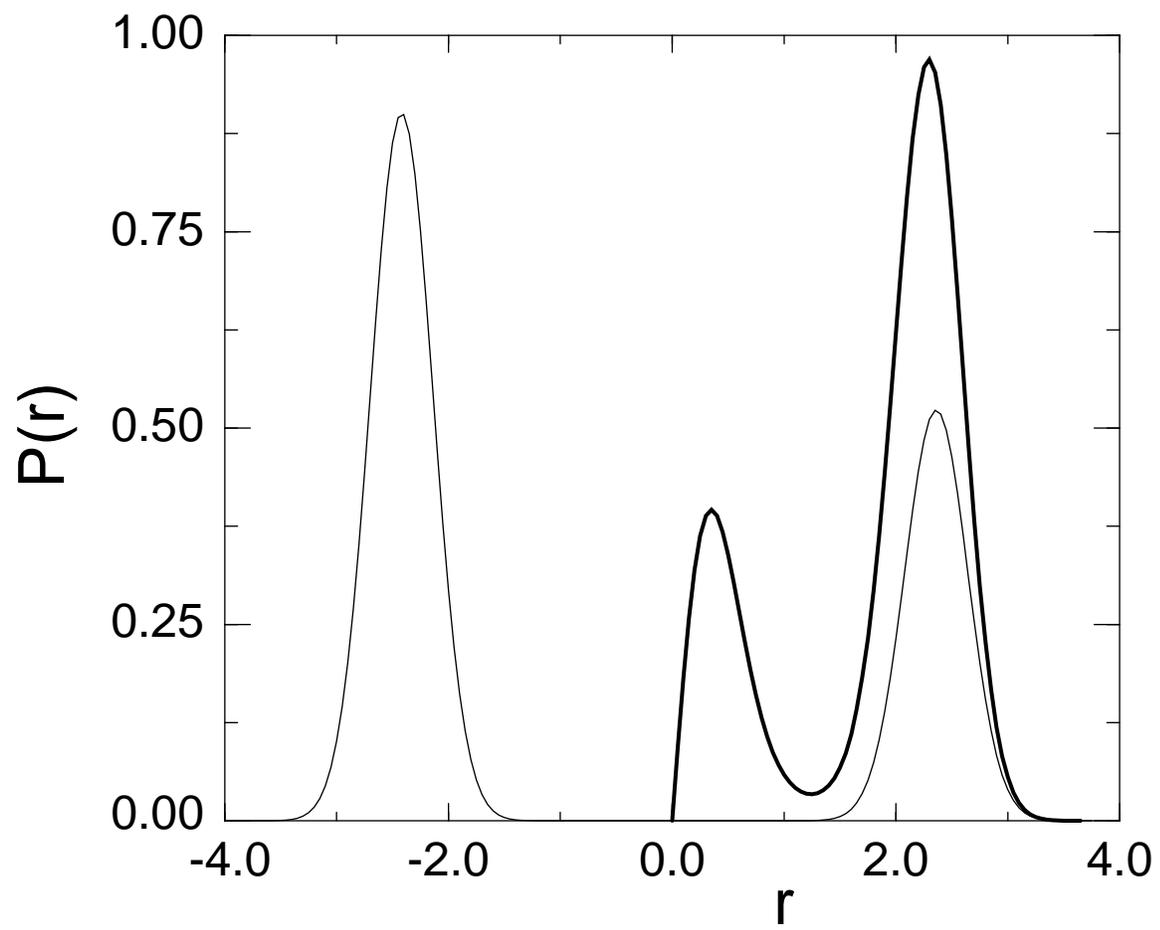
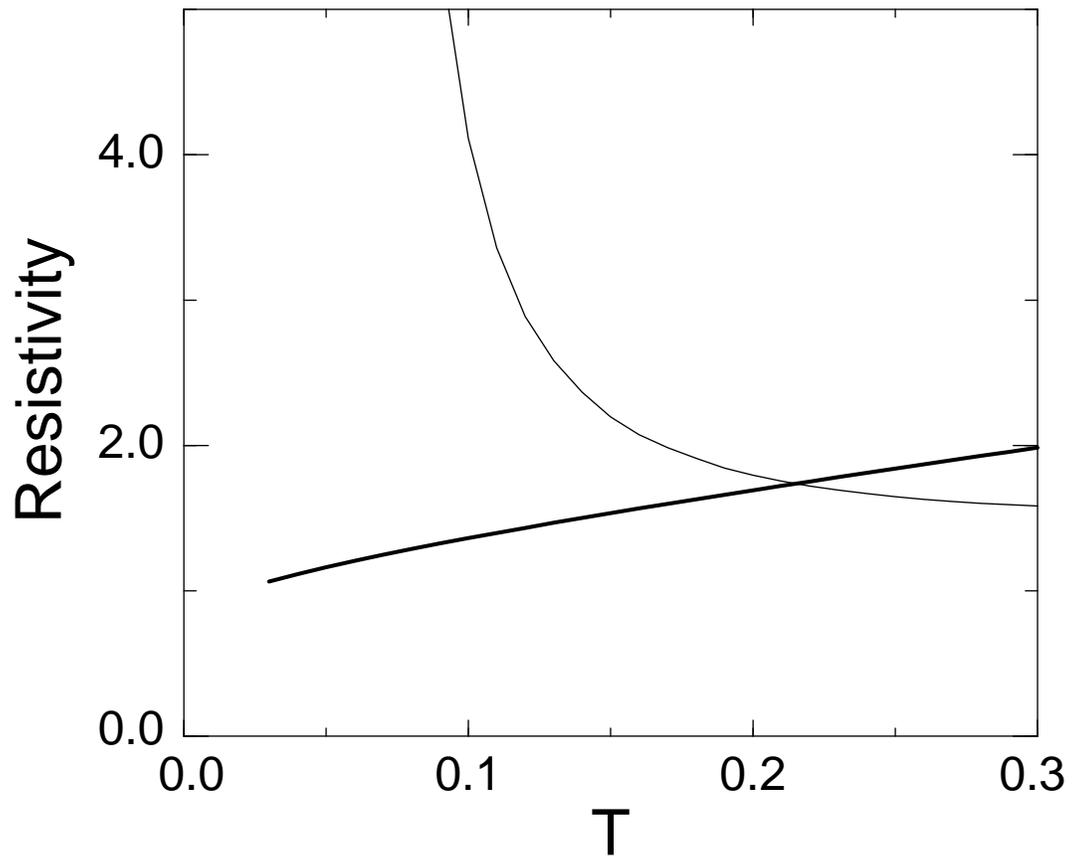
\end{document}